\documentclass[12pt]{iopart}

\usepackage[T1]{fontenc}
\usepackage[utf8]{inputenc}
\expandafter\let\csname equation*\endcsname\relax
\expandafter\let\csname endequation*\endcsname\relax
\usepackage{amsmath,amsthm,mathtools,color,graphicx,upquote}
\usepackage{iopams}
\usepackage{textcomp,gensymb,units}
\usepackage[sc]{mathpazo} % Palatino font
\usepackage{dsfont,bm}
\usepackage[scr=boondoxo,scrscaled=1.05]{mathalfa}
\usepackage[hidelinks]{hyperref}
\usepackage[a4paper]{geometry}
\newcommand{\D}{\mathrm{d}}
\newcommand{\DD}{\operatorname{D}}

\newcommand{\TT}{\mathscr{T}}
\renewcommand{\Pr}{\operatorname{Pr}}
\newcommand{\Cov}{\operatorname{Cov}}
\newcommand{\Var}{\operatorname{Var}}
\DeclareMathOperator{\expectation}{\mathbb{E}}
\newtheorem*{proposition}{Proposition}
\newcommand{\tf}{t_{\mathscr{f}}}

\begin{document}

\title[Classical uncertainty relations and entropy production]{
	Classical uncertainty relations and entropy production in non-equilibrium statistical mechanics 
}

%\date{}

\author{Paolo Muratore-Ginanneschi$^1$ and Luca Peliti$^{2}$ }

\address{$^1$ Department of Mathematics and Statistics, University of Helsinki,	P.O. Box 68, 00014 Helsinki, Finland}

\address{$^2$ Santa Marinella Research Institute, 00058 Santa Marinella, Italy}

\eads{\href{mailto:paolo.muratore-ginanneschi@helsinki.fi}{paolo.muratore-ginanneschi@helsinki.fi},  \href{mailto:luca@peliti.org}{luca@peliti.org}}

\begin{abstract}
	We analyze F\"urth's 1933 classical uncertainty relations in the modern language of stochastic differential equations. Our interest is motivated by applications to non-equilibrium classical statistical mechanics. We show that F\"urth's uncertainty relations are a property enjoyed by  martingales under the measure of a diffusion process. This result implies a lower bound on fluctuations in current velocities of entropic quantifiers of transitions in stochastic thermodynamics. In cases of particular interest, we
	recover an inequality well known in optimal mass transport relating the mean kinetic energy of the current velocity and the squared quadratic Wasserstein distance between the probability distributions of the entropy. In performing our analysis,  we also avail us of an unpublished argument due to Krzysztof Gaw\c{e}dzki to derive a lower bound to the entropy production by transition described by Langevin-Kramers process in terms of the squared quadratic Wasserstein distance between the initial and final states of the transition. Finally, we illustrate how F\"urth's relations admit a straightforward extension to piecewise deterministic processes. We thus show that the results in the paper concern properties enjoyed by general Markov processes. 
\end{abstract}	

% To prevent pagebreak after abstract
{
	\let\newpage\relax    
	\maketitle
}

\section{Introduction}

The notion of uncertainty relations is most often associated with Quantum Mechanics.
This fact is well justified considering the role played by the discovery of the uncertainty relations for the construction of Quantum Mechanics \cite{BuLaMi1996}. Heisenberg observed that  ``\emph{only the uncertainty which is specified by}''  the position momentum uncertainty relation ``\emph{makes room for the validity of the relations which find their most pregnant expression in the quantum-mechanical commutation relations}''~\cite[\S~2]{Hei1927} and concluded that  ``\emph{because all experiments are subject to the laws of quantum mechanics, and therefore to}'' the position momentum uncertainty relation ``\emph{it follows that quantum mechanics establishes the final failure of causality}''~\cite[last sentence]{Hei1927}.

It is interesting to note, however, that already very early in the development of Quantum Theory it was realized that some uncertainty relations do occur also in classical Statistical Physics. Specifically, Reinhold  F\"urth \cite{Fur1933} showed in 1933  that ``\emph{it  is  possible  to  show  that   Heisenberg’s  uncertainty  relations  carry  over  to processes  which  are  governed  by  classical  Statistics  and  that  it  is  thus  possible to  bring  about  new  perspectives  on  the  often  addressed  question  of  the  limit  of measurability with any measurement device}''.
While introducing his result, F\"urth took also care to add  that ``\emph{it might be already known to some}'' in a possible reference to Schr\"odinger's paper \cite{Sch1931} (see also \cite{ChMGSc2021}). We recently offered \cite{PePMG2020} an English translation of F\"urth's paper, originally published in German.
F\"urth's paper played a role for the development of Stochastic Mechanics \cite{Fen1952,Nel1985} and is well known in the community interested in the foundations of Quantum Mechanics (see, e.g., \cite{GhOmRiWe1978}). %\cite{GhOmRiWe1978,GrHaTa1979}. 

Our interest in F\"urth's paper, however, remains rooted in classical statistical physics. Over the last few years, certain types of uncertainty relations, i.e., of universal inequalities connecting the average value of an arbitrary current to its variance and to the average entropy production rate, have recurrently appeared in the context of stochastic  thermodynamics (see, e.g., \cite{BaSe2015,PePi2020} and references therein). 
With this motivation in mind, our aim in the present note is to discuss a general mechanism for uncertainty relations obeyed by certain indicators of Markov processes.

%\marginpar{Revise at the end. One should more definite in the first 4 sections.}
To set the stage for our discussion, in section~\ref{sec:Wiener} we recall some basic concepts in the kinematics of stochastic motion, 
following \cite{Nel1985,Nel2001}. Next, in section~\ref{sec:osm} we derive F\"urth's uncertainty relations in the modern language of stochastic differential equations driven by Wiener processes. Specifically, we consider test functions $ f $ of a process $ \left\{ \bm{\xi}_{t} \right\}_{t\,\geq\,0} $, solution a of stochastic differential equation such that $ f $ is a  martingale (see, e.g., \cite{Kle2005} or \cite{Ste2001}), under the probability measure of the same process. We prove that the variance of $ f $ times the variance of its time-symmetric derivative $ \DD f $ along the paths of $ \left\{ \bm{\xi}_{t} \right\}_{t\,\geq\,0} $ is bounded from below by a strictly positive quantity, known in probability theory as the ``energy of the carr\'e du champ'' defined by $ f $ \cite{BaGeLe2014}.  From the physics stand-point, bounds of this type define uncertainty relations satisfied by open classical systems modeled by stochastic differential equations \cite{ZwaR2001}. We relate F\"urth's uncertainty relations to the so-called ``multivariate inequalities'' known in mathematical statistics~\cite{BiDo2015}.  Finally, we derive an ``integral version'' of the uncertainty relations which we later show to be consequential for stochastic thermodynamics applications.  Section \ref{sec:ada} extends the same result to a larger class of martingales of the solution a of stochastic differential equation. The extension is necessary in order to apply  F\"urth's uncertainty relations to Girsanov's change of measure formula \cite{Kle2005,Ste2001}. Namely, we show in section~\ref{sec:sm}  how F\"urth's uncertainty relations imply a lower bound on the current velocity transporting the probability distribution of in\-form\-ation-theoretic indicators of non-equilibrium processes.  Next, drawing on an unpublished argument due to Krzysztof Gaw\c{e}dzki~\cite{Gaw2021}, we include the inequality just mentioned in a chain of relations implying that current velocity fluctuations during a non-equilibrium transition are bounded from below by the squared quadratic Wasserstein distance~\cite{OtVi2000,Vil2009,PeCu2019} between the probability distributions of the state of the system at the end times of the transition horizon.
Wasserstein distances play a pivotal role in multiple computational  applications of  optimal transport \cite{PeCu2019}.
Finally, in section~\ref{sec:counting}, we derive F\"urth's uncertainty relations for processes solution of stochastic differential equations driven by Poisson processes. These type of processes generically occur when unraveling quantum master equations~\cite{BrPe2002, DoMG2022}. The aim of section~\ref{sec:counting}  is thus to exhibit how to extend the results explicitly derived in the present note to Markov processes of more general form.  

In order to dispel any possible misunderstanding, it is worth emphasizing that we are not trying by any mean to uphold here the equivalence of Quantum Mechanics to a classic stochastic process. On the contrary, we are interested in how phenomena often thought to be exclusively inherent to quantum phenomenology may instead appear in the classical context, in a spirit perhaps closer to F\"urth's idea as well as to the present day investigations \cite{Kir2003,IgToAu2020}.

\section{Osmotic derivative and current velocity} 
\label{sec:Wiener}

We start by recalling some elementary facts about a Markov process 
$ \left\{ \bm{\xi}_{t} \right\}_{t\,\geq\,0} $ generated by the solution of the stochastic differential 
equation driven  by a $ d $-dimensional Wiener process $ \left\{ \bm{w}_{t} \right\}_{t\,\geq\,0} $ .

We suppose that $ \left\{ \bm{\xi}_{t} \right\}_{t\,\geq\,0} $,  is the unique solution of a real valued $ d $-dimensional Itō stochastic differential equation 
\begin{align}
	\label{Wiener:sde}
	\D \bm{\xi}_{t}=\bm{b}(\bm{\xi}_{t},t)\,\D t+\sqrt{\mathds{A}}(\bm{\xi}_{t},t)\,\D \bm{w}_{t},
\end{align}
with the initial condition
\begin{align}
	\label{Wiener:sde2}
	\Pr \left(\bm{x}\,\leq\,\bm{\xi}_{0} \,<\,\bm{x}+\D \bm{x}\right)=\mathscr{p}_{\iota}(\bm{x})\mathrm{d}^{d}\bm{x}.
	\end{align}
Specifically, we suppose that  the drift  $ \bm{b} $ and the diffusion (tensor)  $ \mathds{A} $ fields in (\ref{Wiener:sde}) enjoy regularity properties in $ \mathbb{R}^{d} $ guaranteeing global existence and uniqueness of solutions. As a consequence, $ \left\{ \bm{\xi}_{t} \right\}_{t\,\geq\,0} $ is a Markov process (see, e.g., \S~5.5 of \cite{Kle2005} )  characterized by a transition probability density $ \TT \colon\mathbb{R}^{d}\,\times\,\mathbb{R}_{+}\,\times\,\mathbb{R}^{d}\,\times\,\mathbb{R}_{+}\mapsto \mathbb{R}_{+}$ satisfying for all $ 0\,\leq\,s\,\leq\,u\,\leq\,t $ the Chapman-Kolmogorov equation
\begin{align}
	\TT (\bm{x},t|\bm{y},s)=\int_{\mathbb{R}^{d}}\D^{d}\bm{z}\;
	\TT (\bm{x},t|\bm{z},u)\,\TT (\bm{z},u|\bm{y},s).
	\label{Wiener:CK}
\end{align}
As well known, the Markov property allows us to explicitly compute any multi-time joint probability distribution of $ \left\{ \bm{\xi}_{t} \right\}_{t\,\geq\,0} $ in terms of the transition probability density $ \TT $. In this way we can apply Kolmogorov's extension theorem (see, e.g., \S~2.8 of \cite{Kle2005} ) to define a probability measure $  \operatorname{P}$ on the space of continuous paths $ \mathcal{C}([0,\tf]) $ trodden by sample solutions of (\ref{Wiener:sde})  in any finite time interval $ [0,\tf] $.

The transition probability $ \TT  $, regarded in (\ref{Wiener:CK}) as a function of $ (\bm{x},t) $ satisfies the forward Kolmogorov (Fokker-Planck) equation. As is well known, $ \TT  $ describes a probability density conditional upon the event that the path evolves from $ \bm{y} $  at a time $ s $ earlier than $ t $.
As a function of $ (\bm{y},s) $, $ \TT  $ satisfies  the backward Kolmogorov equation, and yields the mean value of an indicator localized in~$x$ at time~$t$. The latter information is encoded in the definition of the mean forward derivative of any nice test function along the paths of $ \left\{ \bm{\xi}_{t} \right\}_{t\,\geq\,0} $
\begin{align}
	(\DD_{+}f)(\bm{x},t)\coloneqq\lim_{s\,\searrow\,0}\expectation\left(\frac{f(\bm{\xi}_{t+s},t+s)-f(\bm{\xi}_{t},t)}{s}\bigg{|}\bm{\xi}_{t}=\bm{x}\right).
	\label{Wiener:mfd}
\end{align}
A straightforward calculation using the definition of the transition probability density yields the explicit expression of the mean forward derivative
\begin{align}
	(\DD_{+}f)(\bm{x},t)=	\partial_{t}f(\bm{x},t)+(\mathrm{L}f)(\bm{x},t),
	\nonumber
\end{align}
in terms of the generator of the process $ \operatorname{L} $, i.e., the differential operator:
\begin{align}
	(\mathrm{L}f)(\bm{x},t)\coloneqq \left(\left \langle\,\bm{b}(\bm{x},t)\,,\partial_{\bm{x}}\,\right\rangle+ \frac{1}{2}\left \langle\, \mathds{A}(\bm{x},t)\,,\partial_{\bm{x}}\,\otimes\,\partial_{\bm{x}}\,\right\rangle
	\right)f(\bm{x},t).
	\label{Wiener:generator}
\end{align}
Above and in what follows, we use the same symbol $ \left \langle\,\cdot\,,\cdot\,\right\rangle $ to denote the inner product in the case of both vectors and tensors.

 The \textquotedblleft carré du champ\textquotedblright\  operator \cite{BaGeLe2014} acting on test functions $ f $, $ g $ is defined by
\begin{align}
\Gamma\left(f,g\right)\coloneqq \operatorname{L}\left(fg\right)-f \operatorname{L} g - g \operatorname{L} f,
\end{align}
that yields
\begin{align}
\Gamma\left(f,g\right)=\left \langle\,(\bm{\partial}f)(\bm{x},t)\,,\mathds{A}(\bm{x},t)(\bm{\partial}g)(\bm{x},t)\,\right\rangle.
\end{align}
 This operator  specifies the properties of the diffusion tensor from the mean forward derivative:
\begin{align}
\Gamma\left(f,g\right)&=\lim_{s\,\searrow\,0}\expectation\left(\frac{\big{(}f(\bm{\xi}_{t+s},t+s)-f(\bm{\xi}_{t},t)\big{)}\big{(}g(\bm{\xi}_{t+s},t+s)-g(\bm{\xi}_{t},t)\big{)}}{s}\bigg{|}\bm{\xi}_{t}=\bm{x}\right)
\nonumber\\
&
=
	(\DD_{+}f g)(\bm{x},t)-f(\bm{x},t)(\DD_{+}g)(\bm{x},t)-g(\bm{x},t)(\DD_{+}f)(\bm{x},t),
\label{Wiener:cdc}
\end{align}
The mean value of  $\Gamma\left(f,f\right)$  given by
\begin{align}
	\expectation\left\|(\bm{\partial}f)(\bm{\xi}_{t},t)\right\|_{\mathds{A}}^{2}=\expectation\left \langle\,
	(\bm{\partial}f)(\bm{\xi}_{t},t)\,,\mathds{A}(\bm{\xi}_{t},t)(\bm{\partial}f)(\bm{\xi}_{t},t)\,\right\rangle,
%	=\expectation\Big{(}(\operatorname{L}f^{2})(\bm{\xi}_{t},t)-2\,f(\bm{\xi}_{t},t)(\operatorname{L})(\bm{\xi}_{t},t)\Big{)}
\label{Wiener:energy}
\end{align}
will be called the \textit{energy} of the carré du champ.

A further important consequence of the Markov property that we are going to need is Kolmogorov's \emph{time reversal relation} \cite{Kol1937}. The knowledge at any time of the probability density
\begin{align}
	\mathscr{p}(\bm{x},t)=\int_{\mathbb{R}^{d}}\D^{d}\bm{y}\;\TT (\bm{x},t|\bm{y},0)\mathscr{p}_{\iota}(\bm{y}),
	\nonumber
\end{align}
 of $ \left\{ \bm{\xi}_{t} \right\}_{t\,\geq\,0} $,  evolving from a given initial condition (\ref{Wiener:sde2}) together with the transition probability density, specifies the reverse transition probability $ \TT _{R} $ along the path of the probability density:  
\begin{align}
	\TT _{R}(\bm{y},t\big{|}\bm{x},t+s)\mathscr{p}(\bm{x},t+s)=\TT (\bm{x},t+s\big{|}\bm{y},t)\mathscr{p}(\bm{y},t),
	\label{Wiener:Kolmogorov}
\end{align}
where $ s\,\geq\,0 $. The reverse transition probability allows us to define the mean backward derivative \cite{Nel2001} of any test function $ f $:
\begin{align}
(\DD_{-}f)(\bm{x},t)\coloneqq\lim_{s\,\searrow\,0}\expectation\left(\frac{f(\bm{\xi}_{t},t)-f(\bm{\xi}_{t-s},t-s)}{s}\bigg{|}\bm{\xi}_{t}=\bm{x}\right).
\nonumber
\end{align}
According to Nelson's duality relation \cite{Nel2001}, we have 
\begin{align}
(\DD_{-}f)(\bm{x},t)&=\left .\frac{\D }{\D  s}
\int_{\mathds{R}^{d}}\D^{d}y\;\frac{\TT (\bm{x},t\big{|}\bm{y},s)\mathscr{p}(\bm{y},s)}{\mathscr{p}(\bm{x},t)} f(\bm{y},s)\right |_{s=t}
\nonumber\\
&
=(\DD_{+}f)(\bm{x},t)-2\,(\mathfrak{d}f)(\bm{x},t),
\nonumber
\end{align}
where the \textit{osmotic derivative} $\mathfrak{d}f$ is defined by
\begin{align}
(\mathfrak{d}f)(\bm{x},t)\coloneqq \frac{(\DD_{+}f)(\bm{x},t)-(\DD_{-}f)(\bm{x},t)}{2}.
\end{align}
A straightforward computation (see appendix~\ref{ap:osmotic}) yields the explicit expression of the osmotic derivative of the test function $ f $ \cite{Nel2001}:
\begin{align}
	(\mathfrak{d}f)(\bm{x},t)=\frac{1}{2\,\mathscr{p}(\bm{x},t)}\left \langle\,\bm{\partial}_{\bm{x}}\,,\mathds{A}(\bm{x},t)(\bm{\partial}f)(\bm{x},t)\mathscr{p}(\bm{x},t)\,\right\rangle.
	\label{Wiener:osmotic}
\end{align}
Thus, the information contained in the mean forward and backward derivatives is also equivalently encoded in the osmotic derivative and the \textit{current velocity}. The latter quantity is obtained by considering a time symmetric increment:
\begin{align}
	(\DD f)(\bm{x},t)\coloneqq\frac{(\DD_{+}f)(\bm{x},t)+(\DD_{-}f)(\bm{x},t)}{2}
	=(\DD_{+}f)(\bm{x},t)-(\mathfrak{d}f)(\bm{x},t).
	\label{Wiener:cv}
\end{align}
The last expression will be useful in what follows.

\section{Uncertainty relation with respect to the osmotic derivative}
\label{sec:osm}

F\"urth's uncertainty relations consist in the observation that the  product of the variance of any local indicator of the Markov process $ \left\{ \bm{\xi}_{t} \right\}_{t\,\geq\,0} $ with the variance of associated osmotic velocity is bounded from below by the ``energy'' quadratic form induced by the carré du champ operator. The fact is an immediate consequence of the Cauchy-Schwarz inequality. Indeed, under appropriate regularity assumptions we have the following
\begin{proposition}
For any test function $ f\colon\mathbb{R}^{d}\,\times\,\mathbb{R}_{+}\mapsto\mathbb{R} $ one has
	\begin{align}
		\Var \big{(}f(\bm{\xi}_{t},t)\big{)}\,\Var \big{(}\mathfrak{d}f(\bm{\xi}_{t},t)\big{)}
		\,\geq\,
		\frac{1}{4}\left |\expectation\left\|(\bm{\partial}f)(\bm{\xi}_{t},t)\right\|_{\mathds{A}}^{2}\right |^{2},
		\label{osm:ineq}
	\end{align}
\end{proposition}
where in the right-hand side appears the energy of the carré du champ.
\begin{proof}
	Under our assumptions
	\begin{align}
		\expectation(\mathfrak{d}f)(\bm{\xi}_{t},t)=\frac{1}{2}\int_{\mathbb{R}^{d}}\D^{d}\bm{x}\,
		\left \langle\,\bm{\partial}_{\bm{x}}\,,\mathds{A}(\bm{x},t)(\bm{\partial}f)(\bm{x},t)\mathscr{p}(\bm{x},t)\,\right\rangle,
		\nonumber
	\end{align}
vanishes by construction. Thus, by applying the Cauchy-Schwartz inequality, we have
\begin{align}
	&\Var \big{(}f(\bm{\xi}_{t},t)\big{)}\,\Var \big{(}\mathfrak{d}f(\bm{\xi}_{t},t)\big{)}
\nonumber	\\
	&\hspace{0.4cm}\geq\,\left |\int_{\mathbb{R}^{d}}\D^{d}\bm{x}\;\mathscr{p}(\bm{x},t) f(\bm{x},t)\left( \frac{1}{2\,\mathscr{p}(\bm{x},t)}\left \langle\,\bm{\partial}_{\bm{x}}\,,\mathds{A}(\bm{x},t)(\bm{\partial}f)(\bm{x},t)\mathscr{p}(\bm{x},t)\,\right\rangle\right)\right |^{2},
\end{align}
and the result follows after an integration by parts.
\end{proof}
The inequality acquires a particularly suggestive form if $ f $ is a martingale with respect to the probability measure of $ \left\{ \bm{\xi}_{t} \right\}_{t\,\geq\,0} $, i.e., if, for any $ t,s\,\geq\,0 $,
\begin{align}
\expectation\left(f(\bm{\xi}_{t+s},t+s) \big{|} \bm{\xi}_{t}=\bm{x}\right)=f(\bm{x},t).
\nonumber
\end{align}
In such a case, the identity 
\begin{align}
(\DD_{+}f)(\bm{x},t)=0,
\nonumber
\end{align}
holds, and because of (\ref{Wiener:energy}) and (\ref{Wiener:cv}) we can couch (\ref{osm:ineq}) in the form
\begin{align}
	\Var \big{(}f(\bm{\xi}_{t},t)\big{)}\,\Var \big{(}(\DD f)(\bm{\xi}_{t},t)\big{)}
	\,\geq\,
	\frac{1}{4}\left |\expectation\left\|(\bm{\partial}f)(\bm{\xi}_{t},t)\right\|_{\mathds{A}}^{2}\right |^{2}.
	\label{osm:martingale}
\end{align}
We thus see that for martingales the inequality takes the form of an uncertainty relation involving the martingale and, as conjugate variable, its current velocity. 

It is informative to also consider an ``integral version'' of F\"urth's uncertainty relation. To this goal we rewrite (\ref{osm:martingale}) in the form
\begin{align}
	\int_{0}^{t}\mathrm{d}s\,	\Var \big{(}(\DD f)(\bm{\xi}_{s},s)\big{)}
	\,\geq\,
	\int_{0}^{t}\mathrm{d}s\,\frac{\left |\expectation\left\|(\bm{\partial}f)(\bm{\xi}_{s},s)\right\|_{\mathds{A}}^{2}\right |^{2}}{4\,\Var \big{(}f(\bm{\xi}_{s},s)\big{)}}.
	\label{osm:aux}
\end{align}
Next, we observe that the Itō integral representation of the martingale
\begin{align}
	\zeta_{t}=f(\bm{\xi}_{0},0)+\int_{0}^{t}\D s\,\left \langle\,\sqrt{\mathds{A}}(\bm{\xi}_{s},s)\D \bm{w}_{s}\,,(\bm{\partial}f)(\bm{\xi}_{s},s)\,\right\rangle,
	\nonumber
\end{align}
yields the identity
\begin{align}
	\Var f(\bm{\xi}_{t},t)-\Var f(\bm{\xi}_{0},0)=\int_{0}^{t}\D s\,\expectation\left\|(\bm{\partial}f)(\bm{\xi}_{s},s)\right\|_{\mathds{A}}^{2}.
	\label{osm:var}
\end{align}
Thus we can couch (\ref{osm:aux}) into the form
\begin{align}
	\int_{0}^{t}\mathrm{d}s\,	\Var \big{(}(\DD f)(\bm{\xi}_{t},t)\big{)}
	\,\geq\,
	\int_{0}^{t}\mathrm{d}s\,\left(\frac{\mathrm{d}}{\mathrm{d} s}\sqrt{\Var \big{(}f(\bm{\xi}_{s},s)\big{)}}\right)^{2}.
	\nonumber
\end{align}
We can therefore avail us of the Cauchy-Schwarz inequality
\begin{align}
\int_{0}^{t}\mathrm{d}s	\, \dot{\mathscr{x}}_{s}^{2} \,\geq\,\frac{(\mathscr{x}_{t}-\mathscr{x}_{0})^{2}}{t},
	\label{osm:BB}
\end{align}
for
\begin{align}
	\mathscr{x}_{t}=\sqrt{\Var \big{(}f(\bm{\xi}_{t},t)\big{)}},
	\nonumber
\end{align}
to arrive at what may be called ``F\"urth's integral inequality''
\begin{align}
	\int_{0}^{t}\D s\,	\Var \big{(}(\DD f)(\bm{\xi}_{s},s)\big{)}\,\geq\,\frac{\left |\sqrt{\Var f(\bm{\xi}_{t},t)}-\sqrt{\Var f(\bm{\xi}_{0},0)}\right |^{2}}{t}.
	\label{osm:gineq}
\end{align}
In section~\ref{sec:sm} we show that the F\"urth's integral inequality generalizes to martingales the celebrated Benamou-Brenier inequality
widely used in optimal transport theory \cite{BeBr2000}.

\subsection{F\"urth's 1933 example}

F\"urth considered a $ 1d $ Wiener process $ \left\{ w_{t} \right\} _{t\,\geq\,0}$. In such a case
\begin{align}
\DD_{+}x=\expectation w_{t}=0.
\nonumber
\end{align}
F\"urth's uncertainty relation (\ref{osm:martingale}) takes the explicit form
\begin{align}
	(\Var w_{t})\,\Var  \left(\DD w_{t}\right)\geq\,\frac{1}{4}.
	\nonumber
\end{align}
The well known expression of the variance of the Wiener process immediately gives
\begin{align}
	\Var  \left(\DD w_{t}\right)\geq\,\frac{1}{4\,t}.
	\nonumber
\end{align}
In this simple case the inequality is tight. Namely
\begin{align}
	\DD x=-\frac{1}{2}\partial_{x}\ln \mathscr{p}(x,t),
	\nonumber
\end{align}
and therefore
\begin{align}
\Var  \left(\DD w_{t}\right)=\Var \left (\frac{1}{2}\partial_{w_{t}}\ln \mathscr{p}(w_{t},t)\right )=-\frac{1}{4}\expectation\partial_{w_{t}}^{2}\ln\mathscr{p}(w_{t},t)=\frac{1}{4\,t}.
\nonumber
\end{align}
If the initial distribution is Gaussian with variance $ \sigma^{2} $, then 
\begin{align}
\xi_{t}=w_{t}+\eta,
\nonumber
\end{align}
where $ \eta $ is a zero-average Gaussian random variable with variance $ \sigma^{2} $. In such a case (\ref{osm:gineq}) yields a non-trivial lower bound:
\begin{align}
	\int_{0}^{t}\D s\,	\Var (\DD \xi)_{s}=\ln\left(1+\frac{t}{\sigma^{2}}\right)\,\geq\,\left |\sqrt{1+\frac{\sigma^{2}}{t}}-\sqrt{\frac{\sigma^{2}}{t}}\right |^{2}.
	%	\frac{t}{4\,\sigma^{2}\left (1+\frac{t}{2\,\sigma^{2}}\right )}.
	\nonumber
\end{align} 
More generally, harmonic polynomials in any dimension larger than one  are martingales with respect to the Wiener measure~\cite{AxBoRa2001} and therefore satisfy similar inequalities.

\subsection{Relation with multivariate inequalities}

It is worth noticing that Furth's inequality (\ref{osm:ineq}) maybe conceptualized as a dynamics based version  of  the ``multivariate inequalities''  known in mathematical statistics~\cite{Pap1993}. Multivariate inequalities relate the Fisher information matrix of a $ d $-dimensional random variable $ \bm{\xi} $ distributed according to a probability density $ \mathscr{p} $:
\begin{align}
	\mathds{F}=\expectation\Big{(}(\bm{\partial}\ln)\mathscr{p}(\bm{\xi})\,\otimes\,(\bm{\partial}\ln)\mathscr{p}(\bm{\xi})\Big{)},
	\nonumber
\end{align}
to the covariance of a vector-valued indicator $ \bm{g}\,\in\,\mathbb{R}^{n} $ of $ \bm{\xi} $, via the relation
\begin{align}
&	\left \langle\,\bm{v}\,,\Cov(\bm{g}(\bm{\xi}))\,\bm{v}\right\rangle\,\geq
	\nonumber\\
	&\qquad {}\left \langle\,\bm{v}\,,\expectation\Big{(}\bm{g}_{*}(\bm{\xi})\Big{)}\mathds{F}^{-1}\expectation\Big{(}\bm{g}_{*}^{\top}(\bm{\xi})\Big{)}\bm{v}\,\right\rangle,\qquad \mbox{for any }\,\bm{v}\,\in\,\mathbb{R}^{n}.
	\label{osm:multi}
\end{align}
For neatness of notation we used here the push-forward notation (e.g., \cite[\S~$\mathscr{O}$.j]{FraT2012})
\begin{align}
	(\bm{g}_{*})_{\mathscr{i},\mathscr{j}}(\bm{x})=\partial_{x_{\mathscr{j}}}g_{\mathscr{i}}(\bm{x}).
	\nonumber
\end{align}
The proof of multivariate inequalities also follows from an application of the matrix Cauchy-Schwarz inequality (see appendix~\ref{ap:multivariate}). Multivariate inequalities differ from  multivariate Cramér-Rao inequalities (see, e.g., \cite{BiDo2015})  in that differentiation is with respect 
to the value of the random variable and not with respect to parameters of the probability distribution.

\section{F\"urth's inequality for adapted martingales}
\label{sec:ada}

F\"urth's uncertainty relations also extend to any martingale $ \left\{ \zeta_{t} \right\}_{t\,\geq\,0} $ adapted to the natural filtration induced by $ \left\{ \bm{\xi}_{t} \right\}_{t\,\geq\,0} $. Hence, we refer as \textquotedblleft adapted martingales\textquotedblright\ to functionals of $ \left\{ \bm{\xi}_{t} \right\}_{t\,\geq\,0} $  amenable to the Itō integral form
\begin{align}
\zeta_{t}=\zeta_{0}+\int_{0}^{t}\left \langle\,\bm{g}(\bm{\xi}_{s},s)\,,\sqrt{\mathds{A}}^{-1}(\bm{\xi}_{s},s)\D w_{s}\,\right\rangle,
	\label{ada:m}
\end{align}
where $ \zeta_{0} $  is an assigned initial data independent of the Wiener process and $ \bm{g}\colon \mathbb{R}^{d}\,\times\,\mathbb{R}_{+}\mapsto \mathbb{R}^{d}$ is a vector field subject to the integrability condition (see, e.g., the discussion in chapter~6 of \cite{Ste2001})
\begin{align}
	\expectation\left\|\bm{g}(\bm{\xi}_{t},t)\right\|_{\mathds{A}^{-1}}^{2}\,<\,\infty.
	\nonumber
\end{align}
The derivation of the inequality follows immediately by considering the extended Markov process 
\begin{align}
	\left\{ \bm{\chi}_{t} \right\}_{t\,\geq\,0}=\left\{ \bm{\xi}_{t}\,\oplus\,\zeta_{t} \right\}_{t\,\geq\,0}.
	\nonumber
\end{align}
Namely, for any test function $ f\colon\mathbb{R}^{d}\,\times\,\mathbb{R}\,\times\,\mathbb{R}_{+}\mapsto \mathbb{R}$
we get
\begin{align}
&	(\tilde{\DD}_{+}f)(\bm{x},z,t)=
\nonumber\\
&\qquad{} (\DD_{+}f)(\bm{x},z,t)+\frac{\left\|\bm{g}(\bm{x},t)\right\|_{\mathds{A}^{-1}}^{2}}{2}\partial_{z}^{2}f(\bm{x},z,t)
	+\left \langle\,\bm{g}(\bm{x},t)\,,\bm{\partial}_{\bm{x}}\,\right\rangle\partial_{z}f(\bm{x},z,t),
	\nonumber
\end{align}
and
\begin{align}
&	(\tilde{\DD}f)(\bm{x},z,t)=(\tilde{\DD}_{+}f)(\bm{x},z,t)
	-\frac{1}{2\,\mathscr{p}(\bm{x},z,t)}\left \langle\,
	\bm{\partial}_{\bm{x}}\,,\mathds{A}(\bm{x},t)
	(\bm{\partial}f)(\bm{x},z,t)	\mathscr{p}(\bm{x},z,t)\,\right\rangle
	\nonumber\\
&\hspace{0.2cm}	{}-\frac{1}{\mathscr{p}(\bm{x},z,t)}\left \langle\,
	\bm{\partial}_{\bm{x}}\,,\bm{g}(\bm{x},t)
	(\partial_{z}f)(\bm{x},z,t)	\mathscr{p}(\bm{x},z,t)\,\right\rangle
	%\nonumber\\
%&\qquad 
	{}-\frac{\left\|\bm{g}(\bm{x},t)\right\|_{\mathds{A}^{-1}}^{2}}{2\,\mathscr{p}(\bm{x},z,t)}\partial_{z}\Big{(}
	(\partial_{z}f)(\bm{x},z,t)	\mathscr{p}(\bm{x},z,t)\Big{)},
	\nonumber
\end{align}
where 
\begin{align}
	\Pr (\bm{x}\,\leq\,\bm{\xi}_{t}\,<\,\bm{x}+\D^{d}\bm{x} \hspace{0.5cm}\&\hspace{0.5cm} z\,\leq\,\zeta_{t}\,\leq\,z+\D z)=\mathscr{p}(\bm{x},z,t)\,\D^{d}\bm{x}\,\D z.
	\nonumber
\end{align}
In the special case
\begin{align}
	f(\bm{x},z,t)=z,
	\nonumber
\end{align}
we obtain the component of the current velocity along $ \partial_{z} $:
\begin{align}
\tilde{\DD}z	=-\frac{1}{\mathscr{p}(\bm{x},z,t)}\left \langle\,
	\bm{\partial}_{\bm{x}}\,,\bm{g}(\bm{x},t)
		\mathscr{p}(\bm{x},z,t)\,\right\rangle
	-\frac{\left\|\bm{g}(\bm{x},t)\right\|_{\mathds{A}^{-1}}^{2}}{2\,\mathscr{p}(\bm{x},z,t)}\partial_{z}
		\mathscr{p}(\bm{x},z,t).
	\nonumber
\end{align}
The evolution of the probability density of the martingale
\begin{align}
	\Pr (z\,\leq\,\zeta_{t}\,\leq\,z+\D z)=\int_{\mathbb{R}^{d}}\mathscr{p}(\bm{x},z,t)\,\D^{d}\bm{x}\,\D z=\mathscr{m}(z,t)\D z,
	\nonumber
\end{align} 
then obeys the continuity equation
\begin{align}
	\partial_{t}\mathscr{m}(z,t)+\partial_{z}\Big{(}\mathscr{v}(z,t)\mathscr{m}(z,t)\Big{)}=0,
\label{ada:mt}
\end{align}
where 
\begin{align}
	\mathscr{v}(z,t)=-\frac{1}{2\,\mathscr{m}(z,t)}\partial_{z}\int_{\mathbb{R}^{d}}\D^{d}\bm{x}\,\mathscr{p}(\bm{x},z,t)\,\left\|\bm{g}(\bm{x},t)\right\|_{\mathds{A}^{-1}}^{2}.
	\nonumber
\end{align}
Cauchy-Schwartz inequality immediately yields 
\begin{align}
	(\Var \zeta_{t})\Var (\mathscr{v}(\zeta_{t},t))
	\,\geq\,\frac{\left(\expectation\left\|\bm{g}(\bm{\xi}_{t},t)\right\|_{\mathds{A}^{-1}}^{2}\right)^{2}}{4}.
	\label{ada:ineq}
\end{align}

\section{Applications to statistical mechanics}
\label{sec:sm}

Inspection of the explicit expression of the variance of the adapted martingale (\ref{ada:m}) 
\begin{align}
	\Var \zeta_{t}-\Var \zeta_{0}=\int_{0}^{t}\D s\,\expectation\left\|\bm{g}(\bm{\xi}_{s},s)\right\|_{\mathds{A}^{-1}}^{2}
	\label{ada:var}
\end{align}
shows that also in the more general case of adapted martingales the right hand side of (\ref{ada:ineq}) coincides with the square of the speed of increase of the martingale. We are therefore in the position to take the same steps as in the end of section~\ref{sec:osm} and thus to arrive to
the F\"urth's integral inequality
\begin{align}
	\int_{0}^{t}\D s\,\Var (\mathscr{v}(\zeta_{s},s))\,\geq\,\frac{\left |\sqrt{\Var \zeta_{t}}-\sqrt{\Var \zeta_{0}}\right |^{2}}{t}\,.
	\label{sm:ineq}
\end{align}

\subsubsection{Relation with the squared quadratic Wasserstein distance between the initial and the final distribution of the martingale}
\label{sec:BeBr}

Interestingly, the mass conservation equation (\ref{ada:mt}) paves the way for an alternative derivation of (\ref{sm:ineq}) based on ideas coming from optimal transport theory \cite{Vil2009}. The flow $ \Phi $ solution of the deterministic differential equation
\begin{align}
	\dot{\mathscr{z}}_{t}=\mathscr{v}(\mathscr{z}_{t},t)
	\nonumber
\end{align}
specifies the Lagrangian map specifying the solution of (\ref{ada:mt})
\begin{align}
	\mathscr{m}_{t}(z)=\int_{-\infty}^{z}\D w\,\delta(z-\Phi_{t\,s}(w))\mathscr{m}_{s}(w)\,.
	\nonumber
\end{align} 
Hence drawing from Benamou and Brenier \cite{BeBr2000} we observe that the following chain of inequalities holds true
\begin{align}
	&	\int_{0}^{t}\D s\,\expectation\mathscr{v}^{2}(\zeta_{s},s)
	=\int_{0}^{t}\D s\,\int_{\mathbb{R}}\D w\,\mathscr{v}^{2}(\Phi_{s,0}(w),s)\mathscr{m}_{0}(w)
	\nonumber\\
	&\hspace{0.5cm}	\,\geq\,
	\int_{\mathbb{R}}\D w\,\mathscr{m}_{0}(w)  \, t\,\left(\int_{0}^{t}\frac{\D s}{t}\mathscr{v}(\Phi_{s,0}(w),s) \right)^{2}=\frac{1}{t}
	\int_{\mathbb{R}}\D w\,\mathscr{m}_{0}(w) \big{(}\Phi_{s,0}(w)-w\big{)}^{2}\,.
	\nonumber
\end{align} 
We thus obtain
\begin{align}
	\int_{0}^{t}\mathrm{d}s\, \Var (\mathscr{v}(\zeta_{s},s))\,\geq\,\frac{\expectation(\zeta_{t}-\zeta_{0})^{2}}{t}.
		\label{BeBr:Wasserstein}
\end{align}
In words, the time integral over the variance of the of the current velocity is bounded from below by the squared quadratic Wasserstein distance between the probability distributions of the adapted martingale process at evaluated at the boundary times of the integration interval \cite{Vil2009,Gaw2013,PeCu2019}. It is now straightforward to verify that the martingale property 
\begin{align}
	\expectation(\zeta_{t})=\expectation(\zeta_{0}),
	\nonumber
\end{align}
ensures the equivalence of (\ref{sm:ineq}) and (\ref{BeBr:Wasserstein}) whenever $ \zeta_{0} $ is deterministic or in other words when $ \Var\zeta_{0} $ vanishes. This is indeed the case for the statistical mechanics application that we consider below. More generally, a further simple application of Cauchy-Schwarz combined with the observation that the variance of the adapted martingale is monotonically increasing proves that (\ref{sm:ineq}) yields a lower bound for (\ref{BeBr:Wasserstein}) whose evaluation only requires single time probability distributions of the adapted martingale process.

\subsection{Lower bound on fluctuations of the relative entropy flow}
\label{sec:KL}

We denote by  $ \operatorname{Q} $ the path measure generated by the solution of the Itō stochastic differential equation
\begin{subequations}
	\label{KL:sde}
	\begin{align}
		&\label{KL:sde1}
		\D \bm{\xi}_{t}=\Big{(}\bm{b}(\bm{\xi}_{t},t)+\bm{g}(\bm{\xi}_{t},t)\Big{)}\D t+\sqrt{\mathds{A}}(\bm{\xi}_{t},t)\D \bm{w}_{t},
		\\
		&\label{KL:sde2}
		\Pr \left(\bm{x}\,\leq\,\bm{\xi}_{0} \,<\,\bm{x}+\D \bm{x}\right)=\mathscr{p}_{\iota}(\bm{x}).
	\end{align}
\end{subequations}
Upon recalling Girsanov's formula (see, e.g., \cite{Kle2005}) 
 \begin{align}
 	\frac{\D \operatorname{Q}_{[0,t]}}{\D \operatorname{P}_{[0,t]}}=\exp\left(\int_{0}^{t}\left \langle\,\bm{g}(\bm{\xi}_{s},s)\,,\sqrt{\mathds{A}}^{-1}(\bm{\xi}_{s},s)\D w_{s}\,\right\rangle
 	-\frac{1}{2}\int_{0}^{t}\mathrm{d}s\,\left\|\bm{g}(\bm{\xi}_{s},s)\right\|_{\mathds{A}^{-1}}^{2}
 	\right),
 	\nonumber
 \end{align}
 we see that (\ref{ada:m}) for $ \zeta_{0} $ describes fluctuations of the logarithm of the Radon-Nikodym derivative of the path measure $ \operatorname{Q} $ with respect to that $ \operatorname{P} $ generated by (\ref{Wiener:sde}):
\begin{align}
	\zeta_{t}= \ln \frac{\D \operatorname{Q}_{[0,t]}}{\D \operatorname{P}_{[0,t]}}-\expectation\ln \frac{\D \operatorname{Q}_{[0,t]}}{\D \operatorname{P}_{[0,t]}}=\int_{0}^{t}\left \langle\,\bm{g}(\bm{\xi}_{s},s)\,,\sqrt{\mathds{A}}^{-1}(\bm{\xi}_{s},s)\D w_{s}\,\right\rangle.
	\nonumber
\end{align}
Furthermore, the Kullback-Leibler divergence, i.e., the relative entropy between the two measures is
\begin{align}
	\operatorname{K}(\operatorname{P}_{[0,t]}\|\operatorname{Q}_{[0,t]})
	=-\expectation\ln \frac{\D \operatorname{Q}_{[0,t]}}{\D \operatorname{P}_{[0,t]}}
	=\frac{1}{2}\Var \zeta_{t}\,.
	\nonumber
\end{align}
In both the last two expressions $ \expectation $ is as always in this text the expectation value with respect to the path measure $ \operatorname{P} $. %\marginpar{The path measure $\operatorname{P}$ should be introduced in \S~2} 
Thus,  we conclude that (\ref{sm:ineq}) is equivalent to a lower bound on the time average of the fluctuations in the relative entropy current velocity:
\begin{align}
	\overline{\Var (\mathscr{v}(\zeta_{t},t))}:=\frac{1}{t}\int_{0}^{t}\D s\,\Var (\mathscr{v}(\zeta_{s},s))\,\geq\,
	\frac{2\,\operatorname{K}(\operatorname{P}_{[0,t]}\|\operatorname{Q}_{[0,t]})}{t^{2}}\,.
	\label{ada:bound}
\end{align}

\subsection{Lower bound on fluctuations of the entropy production flow}

In stochastic thermodynamics, the evolution of classical open systems is commonly described by stochastic differential equations \cite{PePi2020}.
When both inertial and dissipative effects are included, the open system is naturally described in Darboux coordinates, i.e.,  $ d=2\,n $, and the drift field in (\ref{Wiener:sde}) is of the difference of a Hamiltonian and a gradient vector field
\begin{align}
	\bm{b}(\bm{x},t)=\mathds{J}(\bm{\partial}H)(\bm{x},t)-\mathds{G}(\bm{\partial}H)(\bm{x},t).
	\label{ep:drift}
\end{align}
In (\ref{ep:drift}) we stipulate that $ \mathds{J} $ is a constant antisymmetric matrix characterizing the Hamiltonian vector field, $ \mathds{G} $ is also constant and satisfies Einstein's relation
\begin{align}
	\mathds{G}=\frac{\beta}{2}\mathds{A},
	\nonumber
\end{align}
 for $ \beta $ a positive scalar physically interpreted as inverse temperature.  Finally, the scalar function $ H $ denotes a confining Hamiltonian. Under these hypotheses, the work performed on the open system in the interval $ [0,t] $ is a fluctuating quantity given by
\begin{align}
	\mathcal{W}_{t,0}=\int_{0}^{t}\mathrm{d}s\,\partial_{s}H(\bm{\xi}_{s},s),
	\nonumber
\end{align} 
whilst the heat transferred from the system to the environment modeled by thermal noise is the stochastic integral in Stratonovich sense \cite{SekK1998}
\begin{align}
	\mathcal{Q}_{t,0}=-\int_{0}^{t}\left \langle\,\mathrm{d}\bm{\xi}_{s}\,\overset{\diamond}{,}\,(\bm{\partial}H)(\bm{\xi}_{s},s)\,\right\rangle.
	\nonumber
\end{align}
These definitions permit to identify the first law of thermodynamics as the pathwise identity 
\begin{align}
	H(\bm{\xi}_{t},t)-H(\bm{\xi}_{0},0)=\mathcal{W}_{t,0}-\mathcal{Q}_{t,0}.
	\nonumber
\end{align}
The positive definite quantity obtained by adding the variation of the Gibbs-Shannon entropy of the system and that of the environment identified with the mean heat release \cite{ChGa2008,Gaw2013}
\begin{align}
	0\,\leq\,\expectation \ln \frac{\mathscr{p}(\bm{\xi}_{0},0)}{\mathscr{p}(\bm{\xi}_{t},t)}+\beta\expectation \mathcal{Q}_{t,0}
=\beta\int_{0}^{t}\D s\,\expectation\left\|\bm{\partial}_{\bm{\xi}_{s}}\Big{(}H(\bm{\xi}_{s},s)+\frac{1}{\beta}\ln \mathscr{p}(\bm{\xi}_{s},s)\Big{)}\right\|_{\mathds{G}}^{2},
	\label{ep:def}
\end{align}
is interpreted as the mean entropy production and thus embodies the second law of thermodynamics. Furthermore, inspection of (\ref{ep:def}) also shows that the mean entropy production is amenable to the form (\ref{ada:var}) if we choose
\begin{align}
	\bm{g}(\bm{x},t)=\sqrt{2\,\beta}\,\mathds{G}\,\bm{\partial}_{\bm{x}}\Big{(}H(\bm{x},t)+\frac{1}{\beta}\ln \mathscr{p}(\bm{x},t)\Big{)}.
	\nonumber
\end{align}
We justify this observation in appendix~\ref{ap:ep}. Upon repeating the same steps leading to (\ref{ada:bound}) we conclude that F\"urth's uncertainty relations bring about a lower bound on the variance of the current velocity conjugated to the entropy production  in terms of the entropy production itself:
\begin{align}
\int_{0}^{t}\D s\,\expectation\mathscr{v}^{2}(\zeta_{s},s)\,\geq\,
%\frac{\expectation\zeta_{t}^{2}}{t}=
\frac{\beta}{t}\int_{0}^{t}\D s\,\expectation\left\|\bm{\partial}_{\bm{\xi}_{s}}\Big{(}H(\bm{\xi}_{s},s)+\frac{1}{\beta}\ln \mathscr{p}(\bm{\xi}_{s},s)\Big{)}\right\|_{\mathds{G}}^{2}\,.
	\label{ep:Wasserstein}
\end{align}

\subsubsection{Relation with the squared quadratic Wasserstein distance between the initial and final probability distributions of a state in stochastic thermodynamics}
\label{sec:Gawedzki}

The Hamiltonian $ H $ function in (\ref{ep:drift}) specifies the logarithm of the accompanying distribution
\cite{HaTh1982} of the process $ \left\{ \bm{\xi}_{t} \right\}_{t\,\geq\,0} $
\begin{align}
	\mathscr{p}_{\mathscr{a}}(\bm{x},t)=\frac{e^{-\beta\,H(\bm{x},t)}}{Z_{t}},
	\label{Gawedzki:accomp}
\end{align}
characterized by the condition
\begin{align}
	(\operatorname{L}^{\dagger}\mathscr{p}_{\mathscr{a}})(\bm{x},t)=0\,.
	\nonumber
\end{align}
The hypothesis of a confining Hamiltonian guarantees that the partition function $ Z_{t} $ is finite.
Clearly, the accompanying distribution coincides with the equilibrium distribution if the Hamiltonian is time-independent. % \alert{autonomous}.
In terms of the accompanying distribution (\ref{Gawedzki:accomp}), the entropy production by a classical open system in stochastic thermodynamics is then amenable to the form
\begin{align}
	\mathscr{E}=\frac{1}{\beta}\,\int_{0}^{t}\D s\,\expectation\left\|\sqrt{\mathds{G}}\bm{\partial}_{\bm{\xi}_{s}}\ln \frac{\mathscr{p}(\bm{\xi}_{s},s)}{\mathscr{p}_{\mathscr{a}}(\bm{\xi}_{s},s)}\right\|^{2}.
	\label{ep:ep}
\end{align}
This expression is reminiscent of a time integral over the relative Fisher information \cite{CoTh2006} between the density and the accompanying distribution of the process $ \left\{ \bm{\xi}_{t} \right\}_{t\,\geq\,0} $. 
Under rather general conditions, Otto and Villani proved the existence of logarithmic Sobolev inequalities providing lower bounds to the relative Fisher information in terms of the squared quadratic Wasserstein distance \cite{OtVi2000}. It is thus instructive to derive a similar result 
for $ \expectation\zeta_{t}^{2} $ now regarded as entropy production in stochastic thermodynamics. 
To do so, we adapt an idea of Krzysztof Gaw\c{e}dzki \cite{Gaw2021} to simplify the derivation of lower bounds to the entropy production by transitions obeying a Langevin-Kramers dynamics \cite{PMG2014,PMGSc2014}.

We assume a kinetic plus potential form for the Hamiltonian
\begin{align}
	H(\bm{q},\bm{p},t)=\frac{\left\|\bm{p}\right\|^{2}}{2\,m}+U(\bm{q},t)\,.
	\label{Gawedzki:H}
\end{align}
upon using the Darboux coordinates $ \bm{x}= (\bm{q},\bm{p})$ and write
\begin{align}
	\mathds{J}=
	\begin{bmatrix}
		\mathsf{0} & \mathsf{1}_{d}
		\\
		-\mathsf{1}_{d} & \mathsf{0}
	\end{bmatrix}
	\hspace{1.0cm}\&\hspace{1.0cm}
	\mathds{G}=
	\begin{bmatrix}
		\dfrac{\varepsilon\,\tau}{m}\,\mathsf{1}_{d} & \mathsf{0}
		\\
		\mathsf{0}  & \dfrac{m}{\tau}\mathsf{1}_{d}
	\end{bmatrix}\,.
\nonumber
\end{align}
Note that for $ \varepsilon=0 $ we recover the standard Kramers 1940 model \cite{Kra1940} whereas for $ \varepsilon >0$ environment degrees of freedom modeled by independent Wiener processes directly affect the system position coordinates. For all $ \varepsilon $ the system admits a Boltzmann equilibrium if the mechanical potential in (\ref{Gawedzki:H}) is time independent:
\begin{align}
	U(\bm{q},t)\,\equiv\,U(\bm{q})\,.
	\nonumber
\end{align}

Our goal is to express the entropy production in terms of the current velocity governing the macroscopic -- position coordinate -- probability distribution in order to then apply the Benamou-Brenier inequality.
The starting point is the expression of the Fokker--Planck equation in phase space as a mass continuity equation governed by the current velocity 
\begin{align}
	\partial_{t}\mathscr{p}(\bm{x},t)+\left \langle\,\bm{\partial}_{\bm{x}}\,,\bm{v}(\bm{x},t)\mathscr{p}(\bm{x},t)
	\,\right\rangle=0,
	\nonumber
\end{align}
where
\begin{align}
	\bm{v}(\bm{x},t)=\begin{cases}
		\bm{\partial}_{\bm{p}}H(\bm{x},t)-\dfrac{\varepsilon\,\tau}{m}	\bm{\partial}_{\bm{q}}\left(H(\bm{x},t)+\dfrac{1}{\beta}\ln\mathscr{p}(\bm{x},t) \right) :=\bm{v}^{(q)}(\bm{x},t),
		\\[0.3cm]
		-\bm{\partial}_{\bm{q}}H(\bm{x},t)-\dfrac{m}{\tau}	\bm{\partial}_{\bm{p}}\left(H(\bm{x},t)+\dfrac{1}{\beta}\ln\mathscr{p}(\bm{x},t) \right) :=\bm{v}^{(p)}(\bm{x},t).
	\end{cases}
	\nonumber
\end{align}
The macroscopic probability distribution $ \tilde{\mathscr{p}} $ is the marginal of $ \mathscr{p} $ obtained by integrating out momenta
\begin{align}
	\tilde{\mathscr{p}}(\bm{q},t)=\int_{\mathbb{R}^{d}}\D^{d}\bm{p}\,\mathscr{p}(\bm{q},\bm{p},t)\,.
	\label{Gawedzki:marginal}
\end{align}
It is straightforward to verify that also the macroscopic probability distribution obeys a mass continuity equation
\begin{align}
	\partial_{t}\tilde{\mathscr{p}}(\bm{q},t)+\left \langle\,\bm{\partial}_{\bm{q}}\,,\tilde{\mathscr{p}}(\bm{q},t)\tilde{\bm{v}}(\bm{q},t)\,\right\rangle=0,
	\nonumber
\end{align}
now governed by the macroscopic current velocity specified by the identity
	\begin{align}
	\tilde{	\mathscr{p}}(\bm{q},t)\tilde{\bm{v}}(\bm{q},t)=\int_{\mathbb{R}^{d}}\D^{d}\bm{p}\,\mathscr{p}(\bm{q},\bm{p},t)\,\bm{v}_{t}^{(q)}(\bm{q},\bm{p},t)=
		\tilde{\mathscr{p}}(\bm{q},t)\Big{(}\bm{h}_{t}(\bm{q},t)-\varepsilon \,\bm{k}(\bm{q},t)\Big{)},
		\nonumber
	\end{align}
	where
	\begin{align}
		&	\tilde{\mathscr{p}}(\bm{q},t)\,\bm{h}(\bm{q},t)=\int_{\mathbb{R}^{d}}\D^{d}\bm{p}\,\mathscr{p}(\bm{q},\bm{p},t)\,\bm{\partial}_{\bm{p}}H(\bm{q},\bm{p},t),
		\nonumber\\
		&	\tilde{\mathscr{p}}(\bm{q},t)\,\bm{k}(\bm{q},t)=\frac{\tau}{m}\int_{\mathbb{R}^{d}}\D^{d}\bm{p}\,\mathscr{p}(\bm{q},\bm{p},t)\,\bm{\partial}_{\bm{q}}H(\bm{q},\bm{p},t)-\frac{\tau}{m\,\beta}\bm{\partial}_{\bm{q}}\ln \tilde{\mathscr{p}}(\bm{q},t).
		\nonumber
	\end{align}
Next, we notice that the entropy production, now amenable to the form
\begin{align}
	\mathscr{E}&=		\frac{m\,\beta}{\tau}
	\int_{0}^{\tf}\D t\int_{\mathbb{R}^{2d}}\D^{2d}\bm{x}\;
	\mathscr{p}(\bm{x},t)
	\left( \frac{\varepsilon\,\tau^{2}}{m^{2}}\left\|\bm{\partial}_{\bm{q}}\left(H(\bm{x},t)+\frac{1}{\beta}\ln \mathscr{p}(\bm{x},t)\right)\right\|^{2}\right.\nonumber\\
	&\qquad\qquad \left.{}+\left\|\bm{\partial}_{\bm{p}}\left(H(\bm{x},t)+\frac{1}{\beta}\ln \mathscr{p}(\bm{x},t)\right)\right\|^{2}\right),
	\nonumber
\end{align}
admits the  lower bound 
\begin{align}
	\mathscr{E}\,\geq\,\frac{m\,\beta}{\tau}\,\int_{0}^{\tf}\D t\int_{\mathbb{R}^{d}}\D^{d}\bm{q}\;
	\tilde{\mathscr{p}}(\bm{q},t)\,\left( \varepsilon\,\left\|\bm{k}(\bm{q},t)\right\|^{2}
	+\left\|\bm{h}(\bm{q},t)\right\|^{2}\right)\,.
	\nonumber
\end{align}
To check the inequality it suffices to add and subtract the $ d $-dimensional vector fields $ \bm{k} $ and $ \bm{h} $
into the arguments of, respectively, the first and the second Euclidean $ d $-dimensional squared norm in the right hand side. The inequality then follows by unfolding the squared norms into inner products, neglecting positive definite addends and noticing that cross terms vanish in consequence of the definition of $ \bm{k} $ and $ \bm{h} $.

It now remains to exhibit the relation of the lower bound with the macroscopic current velocity. We take the last step in this direction by observing that for all $ \varepsilon $ larger than zero
\begin{align}
		\left\|\tilde{\bm{v}}(\bm{q},t)\right\|^{2}&=	\left\|\bm{h}(\bm{q},t)-\varepsilon\, \bm{k}(\bm{q},t)\right\|^{2}\nonumber\\
	&=\left\|\bm{h}(\bm{q},t)\right\|^{2}+\varepsilon^{2}\left\|\bm{k}(\bm{q},t)\right\|^{2}-2\,\varepsilon\,\left \langle\,\bm{k}(\bm{q},t)\,,\bm{h}(\bm{q},t)\,\right\rangle
	\nonumber\\
	&	\,\leq\,\varepsilon\left(1+\varepsilon\right)\left\|\bm{k}(\bm{q},t)\right\|^{2}+\left(1+\varepsilon\right)\left\|\bm{h}(\bm{q},t)\right\|^{2}\nonumber\\
	&=(1+\varepsilon)\,\left(\varepsilon\,\left\|\bm{k}(\bm{q},t)\right\|^{2}+\left\|\bm{h}(\bm{q},t)\right\|^{2}\right)\,.
	\nonumber
\end{align}
We thus arrive at
\begin{align}
	&		\mathscr{E}\,\geq\,	\frac{m\,\beta}{\tau}\,\int_{0}^{\tf}\D t \int_{\mathbb{R}^{d}}\D^{d}\bm{q}\,\tilde{\mathscr{p}}(\bm{q},t)\, \frac{\left\|\tilde{\bm{v}}(\bm{q},t)\right\|^{2}}{1+\epsilon}\,.
	\label{ep:bound}
\end{align}
To the expression on the right hand side of (\ref{ep:bound}) we can now apply the Benamou-Brenier  inequality \cite{BeBr2000}  holding this time for the Lagrangian map of the macroscopic density
\begin{align}
	\int_{0}^{t}\D s\,\expectation\|\tilde{\bm{v}}(\bm{\mathscr{q}}_{s},s)\|^{2}\,&=
	\expectation \frac{1}{t}\left(\int_{0}^{t}\D s\,\right)\left(\int_{0}^{t}\D s\,\|\tilde{\bm{v}}(\bm{\mathscr{q}}_{s},s)\|^{2}
	\right)
	\nonumber\\
	&	\geq\,
	\expectation \frac{1}{t}\|\int_{0}^{t}\D s\,\overset{\scriptscriptstyle{\bullet}}{\bm{\mathscr{q}}}_{s}\|^{2}
	=\,\frac{\expectation\left\|\bm{\mathscr{q}}_{t}-\bm{\mathscr{q}}_{0}\right\|^{2}}{t},
	\nonumber
\end{align}
where $ \left\{ \bm{\mathscr{q}}_{t} \right\}_{t\,\geq\,0} $ denotes the position coordinate components of the phase space process $ \left\{ \bm{\xi}_{t} \right\}_{t\,\geq\,0} $. The upshot is
\begin{align}
	\mathscr{E}\,\geq\,	\frac{m\,\beta}{\tau}\,\frac{\expectation\left\|\bm{\mathscr{q}}_{t}-\bm{\mathscr{q}}_{0}\right\|^{2}}{(1+\varepsilon)\,t},
	\nonumber
\end{align}
where $ \left\{ \bm{\mathscr{q}}_{t} \right\}_{t\,\geq\,0} $ denotes the position coordinate components of the phase space process $ \left\{ \bm{\xi}_{t} \right\}_{t\,\geq\,0} $. The inequality states that the variance of the current velocity driving the probability distribution of the entropy production is bounded from below by the squared quadratic Wasserstein distance between the final and initial macroscopic distributions (\ref{Gawedzki:marginal}) of the open system.

As a final remark we notice that the existence of the squared quadratic Wasserstein distance bound to the entropy production provides a mathematically controlled justification of a universal expression of the efficiency at maximum power attained by nano scale Stirling heat engines \cite{ScSe2008}. Namely,  the existence of the bound (\ref{ep:bound}) in conjunction  with the Benamou-Brenier inequality permit to extend the rigorous results holding for open systems operating at the micro-scale and thus modeled by a Langevin-Smoluchowski dynamics \cite{PMGSc2015} also to nano-mechanical systems for which inertial effect are non negligible.

\section{Uncertainty relation for piecewise deterministic processes} 
\label{sec:counting}

In recent years, Markov processes described by a piecewise deterministic processes have attracted a growing interest
in view of their use to model a wide range of physical phenomena (see, e.g., \cite{MaNeWy2008,SuPoRiMa2011,MGMePe2012} and, in particular, in the context of open quantum systems see, e.g., \cite{BaBe1991, BrPe2002, DoMG2022}).
  
It is thus instructive to rephrase F\"urth's inequalities in the case of piecewise deterministic processes and, as a consequence, to hint how they extend to general Markov process.
We thus suppose that $ \left\{ \bm{\xi}_{t} \right\}_{t\,\geq\,0} $ is the solution of the stochastic differential 
equation driven  by a standard $ d $-dimensional Poisson process $ \left\{ \bm{\nu}_{t} \right\}_{t\,\geq\,0} $ .
\begin{subequations}
	\label{Poisson:sde}
	\begin{align}
	&\label{Poisson:sde1}
	\D \bm{\xi}_{t}=\bm{b}(\bm{\xi}_{t},t)\D t+\mathds{X}(\bm{\xi}_{t},t)\D \bm{\nu}_{t}\,,
	\\
	&\label{Poisson:sde2}
	\Pr \left(\bm{x}\,\leq\,\bm{\xi}_{0} \,<\,\bm{x}+\D \bm{x}\right)=\mathscr{p}_{\iota}(\bm{x})\,.
	\end{align}
\end{subequations}
The components of the Poisson process have independent increments and are characterized by
\begin{align}
&\D \bm{\nu}_{t}\,\otimes\,\D \bm{\nu}_{t}=\mathds{1}_{d}\D \bm{\nu}_{t}\,,
\nonumber\\
&\expectation(\D \bm{\nu}_{t}\big{|}\bm{\nu}_{t})=\bm{r}\D t\,.
\nonumber
\end{align}	
where we suppose $ \bm{r} $ a constant vector. The mean forward derivative is in this case
\begin{align}
(\DD_{+}f)(\bm{x},t)&=\lim_{s\,\searrow\,0}\expectation\left(\frac{f(\bm{\xi}_{t+s},t+s)-f(\bm{\xi}_{t},t)}{s}\bigg{|}\bm{\xi}_{t}
=\bm{x}\right)
\nonumber\\
&=
\left(\partial_{t}+\bm{b}(\bm{x},s)\cdot\partial_{\bm{x}}
\right)f(\bm{x},t)+\bm{T}[f](\bm{x},t)\cdot\bm{r}.
\label{Poisson:mfd}
\end{align}	
In (\ref{Poisson:mfd}) we introduced the vector-valued finite increment operator $ \bm{T}[f] $ whose components are
\begin{align}
	T_{i}[f](\bm{x},t)=f\left (\bm{x}+\mathds{X}(\bm{x},t)\bm{e}_{i},t\right )-f(\bm{x},t),\hspace{1.0cm}i=1,\dots,d,
	\nonumber
\end{align}
where $ \bm{e}_{i} $ denotes the $ i $-th element of the canonical basis of $ \mathbb{R}^{d} $. The master equation governing the evolution of the probability density is then
\begin{align}
\partial_{t}\mathscr{p}(\bm{x},t)+\partial_{\bm{x}}\bm{b}(\bm{x},t)\mathscr{p}(\bm{x},t)=\int_{\mathbb{R}^{d}}\D^{d}y\,\sum_{i=1}^{d}\left(
\delta^{(d)}(\bm{y}+\mathds{X}(\bm{y},t)\bm{e}_{i}-\bm{x})-\delta^{(d)}(\bm{y}-\bm{x})\right) r_{i}\mathscr{p}(\bm{y},t),
\nonumber
\end{align}
or in abridged form
\begin{align}
\partial_{t}\mathscr{p}(\bm{x},t)+\partial_{\bm{x}}\bm{b}(\bm{x},t)\mathscr{p}(\bm{x},t)-\bm{T}^{\dagger}[\mathscr{p}](\bm{x},t)\cdot\bm{r}
=0\,.
\nonumber
\end{align}
The square of the forward increment is also finite
\begin{align}
\lim_{s\,\searrow\,0}\expectation\left(\frac{\big{(}f(\bm{\xi}_{t+s},t+s)-f(\bm{\xi}_{t},t)\big{)}^{2}}{s}\bigg{|}\bm{\xi}_{t}=\bm{x}\right)=
\sum_{i=1}^{d} \big{(}\bm{e}_{i}\cdot\bm{T}[f](\bm{x},t)\big{)}^{2}\,r_{i}\,.
\nonumber
\end{align}

\subsection{Current and osmotic derivatives of a test function}

As in the case of stochastic differential equations driven by a Wiener process, we can avail us of the Markov property to define the time symmetric mean derivative of a test function of the process defined by (\ref{Poisson:sde})
\begin{align}
(\DD f)(\bm{x},t)=\lim_{s\,\searrow\,0}\expectation\left(\frac{f(\bm{\xi}_{t+s},t+s)-f(\bm{\xi}_{t-s},t-s)}{2\,s}\bigg{|}\bm{\xi}_{t}=\bm{x}\right)
\,.
\nonumber
\end{align}
A straightforward calculation using Kolmogorov's time reversal identity (\ref{Wiener:Kolmogorov}) and proceeding as in the case of (\ref{Wiener:sde}) yields the expression of the time symmetric mean derivative
\begin{align}
(\DD f)(\bm{x},t)&=\left(\partial_{t}+\bm{b}(\bm{x},t)\cdot\partial_{\bm{x}}\right)f(\bm{x},t)
\nonumber\\
&+\frac{1}{2}\left(\bm{T}[f](\bm{x},t)\cdot\bm{r}
+\frac{f(\bm{x},t)\bm{T}^{\dagger}[\mathscr{p}](\bm{x},t)\cdot\bm{r}-\bm{T}^{\dagger}[f\,\mathscr{p}](\bm{x},t)\cdot\bm{r}}{\mathscr{p}(\bm{x},t)}\right),
\nonumber
\end{align}
and the the ``osmotic'' derivative
\begin{align}
	(\mathfrak{d}f)(\bm{x},t)=\frac{1}{2}\bm{T}[f](\bm{x},t)\cdot\bm{r}
	-\frac{f(\bm{x},t)\bm{T}^{\dagger}[\mathscr{p}](\bm{x},t)\cdot\bm{r}-\bm{T}^{\dagger}[f\,\mathscr{p}](\bm{x},t)\cdot\bm{r}}{2\,\mathscr{p}(\bm{x},t)}.
	\nonumber
\end{align}

\subsection{F\"urth's uncertainty relation}

F\"urth's uncertainty relation follows from the application of the Cauchy-Schwarz inequality to the product of the variances of a scalar martingale
$ f $ with respect to the measure generated by (\ref{Poisson:sde}) and that of its mean symmetric derivative. The lower bound then requires the evaluation of the expectation value
\begin{align}
&\expectation\big{(}f(\bm{\xi}_{t},t)(\mathfrak{d}f)(\bm{\xi}_{t},t)\big{)}=
-\frac{1}{2}\int_{\mathbb{R}^{d}}\D x\,\mathscr{p}(\bm{x},t)\big{(}\bm{T}[f^{2}](\bm{x},t)\cdot\bm{r}-2\,\bm{T}[f](\bm{x},t)\cdot\bm{r}\,f(\bm{x},t)\big{)}
\nonumber\\
&=
\int_{\mathbb{R}^{d}}\D x\,\mathscr{p}(\bm{x},t)\sum_{i=1}^{d}\big{(} f^{2}(\bm{x}+\mathds{X}\bm{e}_{i},t)-f^{2}(\bm{x},t)-2\,f(\bm{x}+\mathds{X}\bm{e}_{i},t)\,f(\bm{x},t)
+2\,f^{2}(\bm{x},t)
\big{)}r_{i}\,.
\nonumber
\end{align}
We thus arrive at
\begin{align}
	\Var \Big{(}f(\bm{\xi}_{t},t)\Big{)}\Var \big{(}\mathfrak{d}f)(\bm{\xi}_{t},t)\big{)}
	\,\geq\,
	\left(\sum_{i=1}^{d}\int_{\mathbb{R}^{d}}\D x\,\mathscr{p}(\bm{x},t)\left(\bm{T}_{i}[f](\bm{x},t)\right)^{2}r_{i}\right)^{2}\,.
	\nonumber
\end{align}
From here, it is straightforward to see how to proceed in order to derive the same type of results obtained in the case of processes generated by the solution of stochastic differential equations driven by Wiener processes.

\section{Conclusions}

Over the last few years, universal inequalities establishing a lower bound for the variance of an empirical current $ \mathscr{J} $ in terms of its square average and the net average entropy production $ \sigma $ 
\begin{align}
	\Var \mathscr{J} \,\geq\, \frac{2\,(\expectation\mathscr{J})^{2}}{e^{\sigma}-1},
	\label{con:gtur}
\end{align}
have attracted considerable attention (see, e.g., \cite{PePi2020,PaSaSeAg2020} and refs therein). In the form written above the inequality is often referred to as ``generalized thermodynamic uncertainty relation'' to discriminate from a ``specialized'' version of the inequality. This latter one is formally obtained by retaining in the right hand side only the leading term of the expansion of the denominator in powers of the net average entropy production.
The derivation of (\ref{con:gtur}) surmises that the joint probability distribution of the entropy production and the current satisfy a detailed fluctuation relation. The derivation is therefore based on statistical considerations whose validity can be ascertained for certain discrete-state Markov processes.  Remarkably, the specialized version of the inequality admits a first principle yet perturbative derivation for near equilibrium processes \cite{PaSaSeAg2020}.
It is thus instructive to compare these results with the statistical physics implications  of F\"urth's uncertainty relations.

First of all, F\"urth's uncertainty relations arise from modeling classical open system by means of stochastic differential equations. As we have shown, the detailed modeling of how environment degrees of freedom act on the system is not essential for the existence of the inequality provided that the resulting system dynamics is a Markov process. From the point of view of first principle derivations, the use of stochastic differential equations is justified by scaling limits applied in the weak environment system coupling regime \cite{ZwaR2001}. Hence the domain validity essentially overlaps with that of thermodynamic uncertainty relations at least as far as general state-of-the-art mathematically rigorous considerations  are concerned. Distinctions can be nevertheless upheld at a more phenomenological level.   

Secondly, F\"urth's uncertainty relations involve the variance and ``the carré du champ'' of physical indicators. In other words, F\"urth's uncertainty relations are statements concerning (at least) second order cumulants of the statistics. In particular, the current velocity defined in section~\ref{sec:sm} is a zero average quantity identically vanishing at equilibrium. The thermodynamic inequality (\ref{con:gtur}) does not carry any non-trivial information for current velocity fluctuations. It must be also noticed that 
weaker bounds on the time integral of the variance of the current fluctuations are easier to derive than bounds based on the squared quadratic Wasserstein distance. They are, however, less informative because they provide trivial bounds for optical traps modeled by transitions corresponding to the change of the curvature of a symmetric single well mechanical potential. In this sense 
F\"urth's uncertainty relations provide stronger bounds stemming, however, form a detailed description of the dynamics rather than general statistical assumptions.

In the beginning of the influential monograph \cite{BaGeLe2014} exploring analytic, probabilistic and geometric properties of Markov semigroups, the authors ask the ``classical mock definition of analysis'':  %\cite{Cou2016}. 
\begin{quote}
	How far can you go with the Cauchy-Schwarz inequality and integration by parts?
\end{quote}
Indeed, these are the only mathematical tools which we needed in this paper to obtain the bounds presented in this paper.  From the physical point of view, reliance on simple mathematical tools seems to us only an advantage as it hints at robustness of the results.

\section{Acknowledgments}

PMG would like to thank Erik Aurell for sharing his last correspondence with the Krzysztof Gaw\c{e}dzki and for making us aware of Krzysztof's beautiful note \cite{Gaw2021}. The idea for deriving the content of section~\ref{sec:Gawedzki} came just after reading \cite{Gaw2021}.  Krzysztof's too early departure deeply saddened us. He and his work remain a source of inspiration for us.

\appendix

\section*{Appendices}

\section{Proof of the representation of the osmotic derivative}
\label{ap:osmotic}

We use the identities
\begin{align}
	&		\partial_{s}\TT (\bm{x},t\big{|}\bm{y},s)=-(\mathrm{L}\TT )(\bm{x},t\big{|}\bm{y},s)\,,
	\nonumber\\
	&			\partial_{s}\mathscr{p}(\bm{y},s)=(\mathrm{L}^{\dagger}\mathscr{p})(\bm{y},s)\,.
	\nonumber
\end{align}
where $ \operatorname{L} $ is the generator (\ref{Wiener:generator}) and $ \operatorname{L}^{\dagger} $ its adjoint in the space of functions integrable with respect to the transition probability of the Markov process.
We then get
\begin{align}
&	(\DD_{-}f)(\bm{x},t)=\partial_{t}f(\bm{x},t)
	\nonumber\\
&\qquad{}	-\lim_{s\uparrow t}\int_{\mathds{R}^{d}}\D^{d}y\,\left(\frac{(\mathrm{L}\TT )(\bm{x},t\big{|}\bm{y},s)\mathscr{p}(\bm{y},s)}{\mathscr{p}(\bm{x},t)}-\frac{\delta^{(d)}(\bm{x}-\bm{y})(\mathrm{L}^{\dagger}\mathscr{p})(\bm{y},t)}{\mathscr{p}(\bm{x},t)}\right)f(\bm{y},s).
	\nonumber
\end{align}
The explicit evaluation hinges upon the identity
\begin{align}
&	\operatorname{L}_{\bm{x}}^{\dagger}\Big{(}\mathscr{p}(\bm{x},t)f(\bm{x},t)\Big{)}=(\mathrm{L}^{\dagger}\mathscr{p})(\bm{x},t)\,f(\bm{x},t)-\mathscr{p}(\bm{x},t) (\mathrm{L}f)(\bm{x},t)
	\nonumber\\
&\qquad{}	+\left \langle\,\bm{\partial}_{\bm{x}}\,,\mathds{A}(\bm{x},t)(\bm{\partial}f)(\bm{x},t)\mathscr{p}(\bm{x},t)\,\right\rangle,
	\nonumber
\end{align}
so that
\begin{align}
	(\DD_{-}f)(\bm{x},t)=\partial_{t}f(\bm{x},t)+(\mathrm{L}f)(\bm{x},t)-\frac{1}{\mathscr{p}(\bm{x},t)}\left \langle\,\bm{\partial}_{\bm{x}}\,,\mathds{A}(\bm{x},t)(\bm{\partial}f)(\bm{x},t)\mathscr{p}(\bm{x},t)\,\right\rangle.
	\nonumber
\end{align}

\section{Proof of the multivariate inequality}
\label{ap:multivariate}

Let $ \bm{g} $ be a vector-valued centered indicator of the process $ \left\{ \bm{\xi}_{t} \right\}_{t\,\geq\,0} $ defined by (\ref{Wiener:sde}). We embed the covariance matrix
\begin{align}
	\mathds{C}=\Cov\Big{(}\bm{g}(\bm{\xi}_{t},t)\Big{)}\,\equiv\,\expectation \Big{(}\bm{g}(\bm{\xi}_{t},t)\,\otimes\,\bm{g}(\bm{\xi}_{t},t)\Big{)},
	\nonumber
\end{align}
in a larger matrix 
\begin{align}
	\mathds{M}=\begin{bmatrix}
		\mathds{C} & \mathds{B}
		\\
		\mathds{B}^{\top} &\mathds{D}
	\end{bmatrix},
	\nonumber
\end{align}
with blocks
\begin{align}
	\mathds{B}=\expectation\left( \bm{g}(\bm{\xi}_{t},t)\,\otimes\,\frac{1}{\mathscr{p}(\bm{\xi}_{t},t)}\partial_{\bm{\xi}_{t}}\mathds{A}(\bm{x},t)\mathscr{p}(\bm{\xi}_{t},t)\right)
	=-\expectation \Big{(}\bm{g}_{*}(\bm{\xi}_{t},t)\mathds{A}(\bm{x},t)\Big{)},
	\nonumber
\end{align}
and
\begin{align}
	\mathds{D}=\expectation\left(\left( \frac{1}{\mathscr{p}(\bm{\xi}_{t},t)}\partial_{\bm{\xi}_{t}}\mathds{A}(\bm{x},t)\mathscr{p}(\bm{\xi}_{t},t)\right)\,\otimes\,\left(\frac{1}{\mathscr{p}(\bm{\xi}_{t},t)}\partial_{\bm{\xi}_{t}}\mathds{A}(\bm{x},t)\mathscr{p}(\bm{\xi}_{t},t)\right)\right)\,.
	\nonumber
\end{align}
The matrix $ \mathds{M} $ is positive definite by construction: for any vector $ \bm{a}\in \mathbb{R}^{2\,d} $ such that $ \bm{a}=\bm{a}_{1}\oplus \bm{a}_{2} $ and $ \bm{a}_{1},\bm{a}_{2} \in \mathbb{R}^{d}$ we get
\begin{align}
	\left \langle\,\bm{a}\,,\mathds{M}\bm{a}\,\right\rangle&=\expectation \left (\left \langle\,\bm{a}_{1}\,,\bm{g}(\bm{\xi}_{t},t)\,\right\rangle\right )^{2}+2\,\expectation\left  (\left \langle\,\bm{a}_{1}\,,\bm{g}(\bm{\xi}_{t},t) \,\right\rangle\frac{1}{\mathscr{p}(\bm{\xi}_{t},t)}\left \langle\,\bm{a}_{2}\,,\partial_{\bm{\xi}_{t}}\mathds{A}(\bm{x},t)\mathscr{p}(\bm{\xi}_{t},t)\,\right\rangle\right )
	\nonumber\\
	& +\expectation \left (\left \langle\, \bm{a}_{2}\,,\frac{1}{\mathscr{p}(\bm{\xi}_{t},t)}\partial_{\bm{\xi}_{t}}\mathds{A}(\bm{x},t)\mathscr{p}(\bm{\xi}_{t},t)\,\right\rangle\right )^{2}
	\nonumber\\
	&=\expectation \left (\left \langle\, \bm{a}_{1}\,,\bm{g}(\bm{\xi}_{t},t)\,\right\rangle+\left \langle\,\bm{a}_{2}\,,\frac{1}{\mathscr{p}(\bm{\xi}_{t},t)}\partial_{\bm{\xi}_{t}}\mathds{A}(\bm{x},t)\mathscr{p}(\bm{\xi}_{t},t)\,\right\rangle\right )^{2}\,\geq\,0,
	\nonumber
\end{align}
so that
\begin{align}
	0\,\leq\,\left \langle\,\bm{a}_{1}\,,\mathds{C}\bm{a}_{1}\,\right\rangle+2\left \langle\, \bm{a}_{1}\,,\mathds{B}\bm{a}_{2}\,\right\rangle+\left \langle\,\bm{a}_{2}\,,\mathds{D}\bm{a}_{2}\,\right\rangle.
	\nonumber
\end{align}
We now choose
\begin{align}
	\bm{a}_{2}=-\mathds{D}^{-1}\mathds{B}^{\top}\bm{a}_{1},
	\nonumber
\end{align}
and obtain
\begin{align}
	0\,&\leq\,\left \langle\,\bm{a}_{1}\,,\mathds{C}\bm{a}_{1}\,\right\rangle-2 \left \langle\,\bm{a}_{1}\,,\mathds{B}\mathds{D}^{-1}\mathds{B}^{\top}\bm{a}_{1}\,\right\rangle+\left \langle\,\bm{a}_{1}\,,\mathds{B}\mathds{D}^{-1}\mathds{D}\mathds{D}^{-1}\mathds{B}^{\top}\bm{a}_{1}\,\right\rangle
\nonumber\\
&	=\left \langle\,\bm{a}_{1}\,,\mathds{C}\bm{a}_{1}\,\right\rangle-\left \langle\,\bm{a}_{1}\,,\mathds{B}\mathds{D}^{-1}\mathds{B}^{\top}\bm{a}_{1}\,\right\rangle.
	\nonumber
\end{align}
Finally, if  $ \bm{a}_{1} $ are Gaussian random vectors with independent components having each zero mean value and unit variance we arrive at
\begin{align}
	\Tr \mathds{C}\,\geq\,\Tr \left(\mathds{B}\mathds{D}^{-1}\mathds{B}^{\top}\right).
	\nonumber
\end{align}

\section{Interpretation of the entropy production in stochastic thermodynamics}
\label{ap:ep}

The interpretation of (\ref{ep:def}) as the second law of thermodynamics is justified by showing that it provides a divergence between probability measures generated by open system dynamics conjugated by a time-reversal operation \cite{ChGa2008}. In this sense, (\ref{ep:def}) is indeed a measure of the irreversibility of a transition.
To exhibit this property, we proceed as in \cite{PMG2014} under, however, the assumption that $ \mathds{G} $ is a non-degenerate  symmetric matrix rather than a projector. The content of this appendix is a special case of those discussed in \cite{ChGa2008} (see also \cite{LeSp1999,MaReMo2000}).

A physically consistent definition of time reversed dynamics in $ [0,\tf]\subset \mathbb{R}_{+} $ is
\begin{subequations}
	\label{ep:bsde}
	\begin{align}
		&\label{ep:bsde1}
		\D _{-}\widehat{\bm{\xi}}_{t}=\Big{(}\mathds{J}(\bm{\partial}H)(\widehat{\bm{\xi}}_{t},t)+\mathds{G}(\bm{\partial}H)(\widehat{\bm{\xi}}_{t},t)\Big{)}\D t+\sqrt{\frac{2\,\mathds{G}}{\beta}}\D \bm{w}_{t},
		\\
		&\label{ep:bsde2}
		\Pr \left(\bm{x}\,\leq\,\widehat{\bm{\xi}}_{\tf} \,<\,\bm{x}+\D \bm{x}\right)=\mathscr{p}(\bm{x},\tf),
	\end{align}
\end{subequations}
where $ \D _{-} $ denotes the backward differential
\begin{align}
	\D _{-}\widehat{\bm{\xi}}_{t}=\widehat{\bm{\xi}}_{t}-\widehat{\bm{\xi}}_{t-\D t},
	\nonumber
\end{align}
and the ``dissipative'' component of the drift in (\ref{ep:bsde1}) is the opposite of that in (\ref{ep:drift}). In (\ref{ep:bsde1}) we also used the fact that increments of the Wiener process are statistically invariant under time reversal. Let now $ \operatorname{P} $ denote the probability measure generated by the paths of (\ref{ep:drift}) and $ \widehat{\operatorname{P}}  $ that of (\ref{ep:bsde}). Clearly $  \operatorname{P}$ weighs events belonging to a $ \sigma $-algebra (\cite{Kle2005}) increasing forward in time (e.g. from $ t =0$ to $ t=\tf $) whereas $ \widehat{\operatorname{P}} $ is a probability measure on a $ \sigma $-algebra increasing from $ \tf $ \textbf{backward} in time. For this reason it is not possible to directly invoke Girsanov formula to compute the Kullback-Leibler divergence between this probability measures. There is, however, a ``workaround''. Namely we can consider the auxiliary process
\begin{align}
	\mathrm{d}\bm{\gamma}_{t}=\mathds{J}(\bm{\partial}H)(\bm{\gamma}_{t},t)\D t+\sqrt{\frac{2\,\mathds{G}}{\beta}}\D \bm{w}_{t}.
	\label{ap:ep:aux}
\end{align} 
The advantage is that the probability measure $ \operatorname{P}^{(\gamma)} $ of (\ref{ap:ep:aux}) is specified by a bi-stochastic transition probability density. In other words, let $ \operatorname{R} $ a path reversal operation in a given time interval. Under $ R $ we obtain that
\begin{align}
	\operatorname{P}^{(\gamma)}[\bm{\gamma}]=(\operatorname{P}^{(\widehat{\gamma})}\circ \operatorname{R})[\bm{\gamma}],
	\label{ap:ep:measure}
\end{align}
where $ \bm{\gamma} $ denotes a path of (\ref{ap:ep:aux})  and $ \operatorname{P}^{(\widehat{\gamma})} $ is the probability measure generated by the backward stochastic differential equation
\begin{align}
	\D _{-}\widehat{\bm{\gamma}}_{t}=\mathds{J}(\bm{\partial}H)(\widehat{\bm{\gamma}}_{t},t)\D t+\sqrt{\frac{2\,\mathds{G}}{\beta}}\D \bm{w}_{t}.
	\nonumber
\end{align}
Combining the identity (\ref{ap:ep:measure}) with the chain rule property enjoyed by the Radon-Nikodym derivative allows us to write
\begin{align}
	\frac{\mathrm{d} \widehat{\operatorname{P}}\circ R}{\mathrm{d} \operatorname{P}}[\bm{\gamma}]=
	\frac{\mathrm{d} \widehat{\operatorname{P}}\circ R}{\mathrm{d} \operatorname{P}^{(\widehat{\gamma})}\circ R}[\bm{\gamma}]
	\frac{\mathrm{d} \operatorname{P}^{(\gamma)}}{\mathrm{d} \operatorname{P}}[\bm{\gamma}].
	\nonumber
\end{align}
We can then use Girsanov formula for forward and backward processes (see e.g. \S~5 of \cite{MeyP1982}) to evaluate the Radon-Nikodym derivatives on the right hand side as functionals of the process $ \left\{ \bm{\gamma}_{t} \right\}_{t\,\geq\,0} $.
%\begin{align}
%&	\frac{\mathrm{d} \operatorname{P}_{[0,t]}}{\mathrm{d} \operatorname{P}_{[0,t]}^{(\gamma)}}=
%		\exp\int_{0}^{t}\left(\sqrt{\frac{\beta}{2}}\left \langle (\bm{\partial}H)(\bm{\gamma}_{s},s) \,, \sqrt{\mathds{G}}\mathrm{d}\bm{w}_{s}\,\right\rangle -\frac{\beta}{4} \left\|(\bm{\partial}H)(\bm{\gamma}_{s},s)\right\|_{\mathds{G}}^{2}\right)
%		\nonumber\\
%&		\frac{\mathrm{d} \operatorname{P}_{[0,t]}^{\flat}\circ R}{\mathrm{d} \operatorname{P}_{[0,t]}^{(\gamma^{\flat})}\circ R}=
%\exp\int_{0}^{t}\left(\sqrt{\frac{\beta}{2}}\left \langle (\bm{\partial}H)(\bm{\gamma}_{s},s) \,\overset{\diamond \diamond}{,}\, \sqrt{\mathds{G}}\mathrm{d}\bm{w}_{s}\,\right\rangle -\frac{\beta}{4} \left\|(\bm{\partial}H)(\bm{\gamma}_{s},s)\right\|_{\mathds{G}}^{2}\right)
%	\nonumber
%\end{align}
%with $ \left\{ \bm{w}_{t} \right\}_{t\,\geq\,0} $ a Wiener process under $ \operatorname{P}^{(\gamma)} $ and $ \diamond \diamond $ denotes the post-point prescription.
 We thus arrive at
\begin{align}
\mathscr{G}[\bm{\gamma}]=\frac{\mathscr{p}(\bm{\gamma}_{\tf},\tf)}{\mathscr{p}(\bm{\gamma}_{0},0)}	\frac{\mathrm{d} \widehat{\operatorname{P}}_{[0,\tf]}\circ R}{\mathrm{d} \operatorname{P}_{[0,\tf]}}[\bm{\gamma}]
	=\exp\left(\int_{0}^{\tf}\Big{(}\beta\left \langle\,(\bm{\partial}H)(\bm{\gamma}_{t},t)\,\overset{\diamond}{,}\,\mathrm{d}\bm{\gamma}_{t}\right\rangle
+\mathrm{d}\ln \mathscr{p}(\bm{\gamma}_{t},t)	\Big{)}\right).
	\nonumber
\end{align}
Finally, we evaluate the Kullback-Leibler divergence between the forward and the backward measure by a further application of Girsanov formula
\begin{align}
	\operatorname{K}(\operatorname{P}_{[0,\tf]}\|\widehat{\operatorname{P}}_{[0,\tf]}\circ R)=
	-\expectation \ln \mathscr{G}[\bm{\xi}]=
	-\expectation^{\operatorname{P}^{(\gamma)}}\left(
	\frac{\mathrm{d} \operatorname{P}_{[0,\tf]}}{\mathrm{d} \operatorname{P}_{[0,\tf]}^{(\gamma)}}[\bm{\gamma}]\ln \mathscr{G}[\bm{\gamma}]\right).
	\nonumber
\end{align}
In the rightmost expression $ \expectation ^{\operatorname{P}^{(\gamma)}} $ denotes the expectation value with respect to the probability measure $ \operatorname{P}_{[0,\tf]}^{(\gamma)} $. A straightforward calculation yields
\begin{align}
	\operatorname{K}(\operatorname{P}_{[0,\tf]}\|\widehat{\operatorname{P}}_{[0,\tf]}\circ R)=\beta\int_{0}^{\tf}\D s\,\expectation\left\|\bm{\partial}_{\bm{\xi}_{s}}\Big{(}H(\bm{\xi}_{s},s)+\frac{1}{\beta}\ln \mathscr{p}(\bm{\xi}_{s},s)\Big{)}\right\|_{\mathds{G}}^{2}.
	\nonumber
\end{align}
%\begin{align}
%&		\expectation \left(\int_{0}^{t}\left(\beta\left \langle\,(\bm{\partial}H)(\bm{\xi}_{s},s)\,\overset{\diamond}{,}\,\mathrm{d}\bm{\xi}_{s}\right\rangle
%		+\mathrm{d}\ln \mathscr{p}(\bm{\xi}_{s},s)	\right)\right)
%\nonumber\\
%&=	\expectation \int_{0}^{t}\mathrm{d}s\,\beta\left \langle\,\bm{v}(\bm{\xi}_{s},s)\,,(\bm{\partial}H)(\bm{\xi}_{s},s)+\frac{1}{\beta}\bm{\partial}_{\bm{\xi}_{s}}\ln\mathscr{p}(\bm{\xi}_{s},s)\,\right\rangle
%\nonumber\\
%&=	-\beta\expectation \int_{0}^{t}\mathrm{d}s\,\left\|(\bm{\partial}H)(\bm{\xi}_{s},s)+\frac{1}{\beta}\bm{\partial}_{\bm{\xi}_{s}}\mathscr{p}(\bm{\xi}_{s},s)\right\|_{\mathds{G}}^{2}
%	\nonumber
%\end{align}
%having used the definition of current velocity
%\begin{align}
%	\bm{v}(\bm{\xi}_{s},s)=\mathds{J}(\bm{\partial}H)(\bm{\xi}_{s},s)-\mathds{G}(\bm{\partial}H)(\bm{\xi}_{s},s)-\frac{1}{\beta}
%	\mathds{G}\bm{\partial}_{\bm{\xi}_{s}}\ln \mathscr{p}(\bm{\xi}_{s},s)
%	\nonumber
%\end{align}
% and the properties of the Stratonovich integral (see e.g. \cite{Nel1985} \S~5).
 
\section*{References}
 \bibliography{Fuerth_martingales}{}

@Book{BuLaMi1996,
  author    = {Paul Busch and Pekka J. Lahti and Peter Mittelstaedt},
  publisher = {Springer Berlin Heidelberg},
  title     = {{The Quantum Theory of Measurement}},
  year      = {1996},
  edition   = {II},
  isbn      = {978-3-540-37205-9},
  series    = {Lecture Notes in Physics Monographs},
  date      = {1996},
  doi       = {10.1007/978-3-540-37205-9},
  pages     = {XIII,181},
  url       = {https://doi.org/10.1007/978-3-540-37205-9},
}

@Article{Fur1933,
  author    = {Reinhold F\"urth},
  journal   = {Zeitschrift f\"ur Physik},
  title     = {{\"Uber einige Beziehungen zwischen klassischer Statistik und Quantenmechanik}},
  year      = {1933},
  month     = {March},
  number    = {3-4},
  pages     = {143--162},
  volume    = {81},
  doi       = {10.1007/bf01338361},
  publisher = {Springer Nature},
}

@Article{Sch1931,
  author    = {Erwin Schr\"odinger},
  journal   = {Sitzungsberichte der preussischen Akademie der Wissenschaften, physikalische mathematische Klasse},
  title     = {{\"Uber die Umkehrung der Naturgesetze}},
  year      = {1931},
  number    = {9},
  pages     = {144-153},
  volume    = {8},
  doi       = {10.1002/ange.19310443014},
}

@Article{ChMGSc2021,
  author    = {Rapha\"el Chetrite and Paolo Muratore-Ginanneschi and Kay Schwieger},
  journal   = {The European Physical Journal H},
  title     = {{E. Schrödinger's 1931 paper {\textquotedblleft}On the Reversal of the Laws of Nature{\textquotedblright} [{\textquotedblleft}Über die Umkehrung der Naturgesetze{\textquotedblright}, Sitzungsberichte der preussischen Akademie der Wissenschaften, physikalisch-mathematische Klasse, 8 N9 144-153]}},
  year      = {2021},
  month     = {nov},
  number    = {1},
  volume    = {46},
  doi       = {10.1140/epjh/s13129-021-00032-7},
  eprint    = {2105.12617},
  publisher = {Springer Science and Business Media {LLC}},
  url       = {https://doi.org/10.1140%2Fepjh%2Fs13129-021-00032-7},
}

@Article{PePMG2020,
  author        = {Luca Peliti and Paolo Muratore-Ginanneschi},
  journal       = {arXiv:2006.03740 eprint},
  title         = {{R. F\"urth' s 1933 paper "On certain relations between classical Statistics and Quantum Mechanics" ["Über einige Beziehungen zwischen klassischer Statistik und Quantenmechanik", \textit{Zeitschrift für Physik,} \textbf{81} 143-162]}},
  year          = {2020},
  month         = jun,
  archiveprefix = {arXiv},
  eprint        = {2006.03740},
  primaryclass  = {physics.hist-ph},
  url           = {https://doi.org/10.48550/arXiv.2006.03740},
}

@Article{Fen1952,
  author    = {F\'enyes, Imre},
  journal   = {Zeitschrift f\"ur Physik},
  title     = {{Eine wahrscheinlichkeitstheoretische Begr\"undung und Interpretation der Quantenmechanik}},
  year      = {1952},
  issn      = {1434-601X},
  month     = {February},
  number    = {1},
  pages     = {81--106},
  volume    = {132},
  doi       = {10.1007/bf01338578},
  publisher = {Springer Science + Business Media},
}

@Book{Nel1985,
  author    = {Edward Nelson},
  publisher = {Princeton University Press},
  title     = {{Quantum fluctuations}},
  year      = {1985},
  isbn      = {\hspace{-0.1cm}-13: 978-0-691-08379-7},
  series    = {Princeton series in Physics},
  pages     = {146},
  url       = {https://web.math.princeton.edu/~nelson/books.html},
}

@Book{Nel2001,
  author    = {Edward Nelson},
  publisher = {Princeton University Press},
  title     = {{Dynamical Theories of Brownian Motion}},
  year      = {2001},
  edition   = {2nd},
  isbn      = {\hspace{-0.1cm}-13: 978-0-691-07950-9},
  pages     = {148},
  url       = {https://web.math.princeton.edu/~nelson/books.html},
}

@Article{BaSe2015,
  author    = {Andre C. Barato and Udo Seifert},
  journal   = {Physical Review Letters},
  title     = {{Thermodynamic Uncertainty Relation for Biomolecular Processes}},
  year      = {2015},
  month     = {apr},
  number    = {15},
  volume    = {114},
  doi       = {10.1103/physrevlett.114.158101},
  eprint    = {1502.05944},
  publisher = {American Physical Society ({APS})},
  url       = {https://doi.org/10.1103%2Fphysrevlett.114.158101},
}

@Book{PePi2020,
  author    = {Peliti, Luca and Pigolotti, Simone},
  publisher = {Princeton University Press},
  title     = {{Stochastic Thermodynamics}},
  year      = {2021},
  isbn      = {9780691215525},
  url       = {https://press.princeton.edu/books/ebook/9780691215525/stochastic-thermodynamics},
}

@Unpublished{Gaw2021,
  author  = {Krzysztof Gawędzki},
  note    = {Note written for Salambo Dago in July 2020, and communicated by the author to Erik Aurell},
  title   = {{Improved 2nd Law of Stochastic Thermodynamics for underdamped Langevin process}},
  month   = sep,
  year    = {2021},
  comment = {Från: Krzysztof Gawedzki [krzysztof.gawedzki@gmail.com] Skickat: den 16 september 2021 21:47 Till: Erik Aurell Ämne: Your mail ``Dear Erik, could you resend your last message becauseI have removed it by mistake. As far as the Paolo's work on underdamperd case is concerned, I tried to summarize it in a note for Salambo Dago in July 2020 that I attach here. I am not sure that it is clearer than Paolo's original, but it helped me. Best, Krzysztof''},
}

@Article{OtVi2000,
  author    = {Felix Otto and Cedric Villani},
  journal   = {Journal of Functional Analysis},
  title     = {{Generalization of an inequality by Talagrand and links with the Logarithmic Sobolev Inequality}},
  year      = {2000},
  month     = {jun},
  number    = {2},
  pages     = {361--400},
  volume    = {173},
  doi       = {10.1006/jfan.1999.3557},
  publisher = {Elsevier {BV}},
  url       = {https://doi.org/10.1006%2Fjfan.1999.3557},
}

@Article{SuPoRiMa2011,
  author        = {Suweis, Samir and Porporato, Amilcare and Rinaldo, Andrea and Maritan, Amos},
  journal       = {Physical Review E},
  title         = {{Prescription-induced jump distributions in multiplicative Poisson processes}},
  year          = {2011},
  issn          = {1550-2376},
  month         = {June},
  number        = {6},
  pages         = {061119},
  volume        = {83},
  archiveprefix = {arXiv},
  doi           = {10.1103/physreve.83.061119},
  eprint        = {1207.2332},
  primaryclass  = {physics.geo-ph},
  publisher     = {American Physical Society (APS)},
}

@Book{BaGeLe2014,
  author    = {Bakry, Dominique and Gentil, Ivan and Ledoux, Michel},
  publisher = {Springer International Publishing},
  title     = {{Analysis and Geometry of Markov Diffusion Operators}},
  year      = {2014},
  isbn      = {978-3-319-00227-9},
  note      = {},
  series    = {Grundlehren der mathematischen Wissenschaften},
  doi       = {10.1007/978-3-319-00227-9},
  issn      = {2196-9701},
  journal   = {Grundlehren der mathematischen Wissenschaften},
  pages     = {552},
}

@Article{Kir2003,
  author        = {Kirkpatrick, Kim Allen},
  journal       = {Foundations of Physics Letters},
  title         = {{``Quantal'' behavior in classical probability}},
  year          = {2003},
  month         = jun,
  number        = {3},
  pages         = {199--224},
  volume        = {16},
  archiveprefix = {arXiv},
  doi           = {10.1023/A:1025910725022},
  eprint        = {quant-ph/0106072v6},
  journaltitle  = {Found Phys Lett 16(3), 199-224 (2003)},
  primaryclass  = {quant-ph},
}

@Article{DoMG2022,
	author        = {Brecht Donvil and Paolo Muratore-Ginanneschi},
	journal       = {Nature Communications},
	title         = {{Quantum trajectory framework for general time-local master equations}},
	year          = {2022},
	doi           = {10.1038/s41467-022-31533-8},
	eprint        = {2102.10355},
	pages         = {4140},
	url           = {https://doi.org/10.1038/s41467-022-31533-8},
	volume        = {13},
	abbr          = {Nat. Commun.},
	archiveprefix = {arXiv},
	month         = jul,
	primaryclass  = {quant-ph},
	ranking       = {{Quantum trajectory framework for general time-local master equations}},
}

@Article{BaBe1991,
  author        = {Alberto Barchielli and Viacheslav P. Belavkin},
  journal       = {Journal of Physics A: Mathematical and General},
  title         = {{Measurements continuous in time and a posteriori states in quantum}},
  year          = {1991},
  month         = dec,
  number        = {7},
  volume        = {24},
  archiveprefix = {arXiv},
  doi           = {10.1088/0305-4470/24/7/022},
  eprint        = {quant-ph/0512189v1},
  journaltitle  = {J. Phys. A: Math. Gen. 24 (1991) 1495--1514},
  primaryclass  = {quant-ph},
}

@Article{ScSe2008,
  author        = {Schmiedl, Tim and Seifert, Udo},
  journal       = {EPL (Europhysics Letters)},
  title         = {{Efficiency at maximum power: An analytically solvable model for stochastic heat engines}},
  year          = {2008},
  issn          = {1286-4854},
  month         = {January},
  number        = {2},
  pages         = {20003},
  volume        = {81},
  archiveprefix = {arXiv},
  doi           = {10.1209/0295-5075/81/20003},
  eprint        = {0710.4097},
  primaryclass  = {cond-mat.stat-mech},
  publisher     = {IOP Publishing},
}

@Article{PMGSc2015,
  author        = {Muratore-Ginanneschi, Paolo and Schwieger, Kay},
  journal       = {EPL (Europhysics Letters)},
  title         = {{Efficient protocols for Stirling heat engines at the micro-scale}},
  year          = {2015},
  month         = {October},
  pages         = {20002},
  volume        = {112},
  archiveprefix = {arXiv},
  doi           = {10.1209/0295-5075/112/20002},
  eprint        = {1503.05788},
  primaryclass  = {cond-mat.stat-mech},
}

@Article{Kra1940,
  author    = {Hendrik Anthony Kramers},
  journal   = {Physica},
  title     = {{Brownian motion in a field of force and the diffusion model of chemical reactions}},
  year      = {1940},
  month     = {April},
  number    = {4},
  pages     = {284–304},
  volume    = {7},
  doi       = {10.1016/S0031-8914(40)90098-2},
}

@Book{CoTh2006,
  author    = {Cover, Thomas M. and Thomas, Joy A.},
  publisher = {Wiley-Blackwell},
  title     = {{Elements of Information Theory}},
  year      = {2006},
  edition   = {2},
  isbn      = {978-0-471-24195-9},
  series    = {Telecommunications and Signal Processing},
  pages     = {776},
}

@Article{HaTh1982,
  author    = {Peter H\"{a}nggi and Harry Thomas},
  journal   = {Physics Reports},
  title     = {{Stochastic processes: Time evolution, symmetries and linear response}},
  year      = {1982},
  month     = {August},
  number    = {4},
  pages     = {207--319},
  volume    = {88},
  doi       = {10.1016/0370-1573(82)90045-x},
  publisher = {Elsevier {BV}},
}

@Article{Gaw2013,
  author         = {Gaw\k{e}dzki, Krzysztof},
  journal        = {arXiv:1308.1518},
  title          = {{Fluctuation Relations in Stochastic Thermodynamics}},
  year           = {2013},
  note           = {Lecture notes},
  archiveprefix  = {arXiv},
  comments       = {45 pages, 16 figures},
  eprint         = {1308.1518},
  oai2identifier = {1308.1518},
  primaryclass   = {math-ph},
}

@Article{ChGa2008,
  author        = {Ch{\'e}trite, Rapha\"el and Gaw\k{e}dzki, Krzysztof},
  journal       = {Communications in Mathematical Physics},
  title         = {{Fluctuation relations for diffusion processes}},
  year          = {2008},
  month         = sep,
  number        = {2},
  pages         = {469--518},
  volume        = {282},
  archiveprefix = {arXiv},
  doi           = {10.1007/s00220-008-0502-9},
  eprint        = {0707.2725},
  primaryclass  = {math-ph},
}

@Book{Ste2001,
  author    = {Steele, J. Michael},
  publisher = {Springer},
  title     = {{Stochastic calculus and financial applications}},
  year      = {2001},
  series    = {Applications of mathematics},
  volume    = {45},
  pages     = {300},
}

@Book{BiDo2015,
  author    = {Peter J. Bickel and Kjell A. Doksum},
  publisher = {CRC Press Taylor \& Francis Group},
  title     = {{Mathematical Statistics. Basic Ideas and Selected Topics }},
  year      = {2015},
  edition   = {II},
  series    = {Texts in statistical science},
  volume    = {I},
}

@Article{Pap1993,
  author    = {Papathanasiou, Vassilis},
  journal   = {Journal of Multivariate Analysis},
  title     = {Some Characteristic Properties of the {F}isher Information Matrix via {C}acoullos-Type Inequalities},
  year      = {1993},
  month     = {feb},
  number    = {2},
  pages     = {256--265},
  volume    = {44},
  doi       = {10.1006/jmva.1993.1014},
  publisher = {Elsevier {BV}},
  url       = {https://doi.org/10.1006%2Fjmva.1993.1014},
}

@Book{Kle2005,
  author    = {Fima C. Klebaner},
  publisher = {Imperial College Press},
  title     = {Introduction to stochastic calculus with applications},
  year      = {2005},
  edition   = {2},
  isbn      = {\hspace{-0.1cm}-13: 978-1-86094-555-7},
  pages     = {416},
}

@Book{AxBoRa2001,
  author    = {Axler, Sheldon Jay and Bourdon, Paul and Ramey, Wade},
  publisher = {Springer},
  title     = {{Harmonic function theory}},
  year      = {2001},
  edition   = {2},
  series    = {Graduate texts in mathematics},
  volume    = {137},
  pages     = {259},
}

@Article{Kol1937,
  author      = {Andre\v{i} Nikolaevich Kolmogorov},
  journal     = {Mathematische Annalen},
  title       = {{Zur Umkehrbarkeit der statistischen Naturgesetze}},
  year        = {1937},
  issn        = {0025-5831},
  pages       = {766-772},
  volume      = {113},
  affiliation = {Moskau},
  doi         = {10.1007/BF01571664},
  issue       = {1},
  publisher   = {Springer Berlin / Heidelberg},
}

@Article{IgToAu2020,
  author        = {Rodrigo Iglesias and Fernando Tohm\'e and Marcelo Auday},
  journal       = {Open Systems \& Information Dynamics},
  title         = {{Contextuality scenarios arising from networks of stochastic processes}},
  year          = {2016},
  month         = jun,
  number        = {03},
  pages         = {1650012},
  volume        = {23},
  archiveprefix = {arXiv},
  doi           = {10.1142/S1230161216500128},
  eprint        = {2006.12432v1},
  journaltitle  = {Open Systems \& Information Dynamics, Vol. 23, No. 03, 1650012 (2016)},
  primaryclass  = {quant-ph},
}

@Article{BeBr2000,
  author    = {Benamou, Jean-David and Brenier, Yann},
  journal   = {Numerische Mathematik},
  title     = {{A computational fluid mechanics solution to the Monge-Kantorovich mass transfer problem}},
  year      = {2000},
  issn      = {0945-3245},
  month     = {January},
  number    = {3},
  pages     = {375–393},
  volume    = {84},
  doi       = {10.1007/s002110050002},
  publisher = {Springer Science + Business Media},
}

@Book{Vil2009,
  author    = {Villani, C\'edric},
  publisher = {Springer},
  title     = {{Optimal transport: old and new}},
  year      = {2009},
  series    = {Grundlehren der mathematischen Wissenschaften},
  volume    = {338},
  pages     = {973},
  url       = {http://www.umpa.ens-lyon.fr/~cvillani/surveys.html#oldnew},
}

@Article{PaSaSeAg2020,
  author    = {Soham Pal and Sushant Saryal and Dvira Segal and T. S. Mahesh and Bijay Kumar Agarwalla},
  journal   = {Physical Review Research},
  title     = {{Experimental study of the thermodynamic uncertainty relation}},
  year      = {2020},
  month     = {may},
  number    = {2},
  volume    = {2},
  doi       = {10.1103/physrevresearch.2.022044},
  eprint    = {1912.08391},
  publisher = {American Physical Society ({APS})},
  url       = {https://doi.org/10.1103%2Fphysrevresearch.2.022044},
}

@Article{Hei1927,
  author    = {Werner Heisenberg},
  journal   = {Zeitschrift f\"ur Physik},
  title     = {{\"Uber den anschaulichen Inhalt der quantentheoretischen Kinematik und Mechanik. }},
  year      = {1927},
  pages     = {172},
  volume    = {43},
  doi       = {10.1007/978-3-642-61659-4_30},
  note = {English translation: `The physical content of quantum kinematics and mechanics' in Wheeler, J. A. \& Zurek, W. H. (Eds.) Quantum Theory and Measurement, Princeton University Press, 1983, 811},
  url       = {https://doi.org/10.1007%2F978-3-642-61659-4_30},
}

@Article{GhOmRiWe1978,
  author    = {Giancarlo Ghirardi and C. Omero and Alberto Rimini and Tullio Weber},
  journal   = {Rivista del Nuovo Cimento},
  title     = {{The stochastic interpretation of quantum mechanics: A critical review}},
  year      = {1978},
  month     = {mar},
  number    = {3},
  pages     = {1--34},
  volume    = {1},
  doi       = {10.1007/bf02724445},
  publisher = {Springer Science and Business Media {LLC}},
  url       = {https://doi.org/10.1007%2Fbf02724445},
}

@Book{BrPe2002,
  author    = {Breuer, Heinz-Peter and Petruccione, Francesco},
  publisher = {Oxford University Press},
  title     = {{The Theory of Open Quantum Systems}},
  year      = {2002},
  edition   = {reprint},
  isbn      = {978-0-199-21390-0},
  doi       = {10.1093/acprof:oso/9780199213900.001.0001},
  pages     = {XXII, 636},
}

@Article{PMG2014,
  author        = {Muratore-Ginanneschi, Paolo},
  journal       = {Journal of Statistical Mechanics: Theory and Experiment},
  title         = {{On extremals of the entropy production by “Langevin--Kramers” dynamics}},
  year          = {2014},
  issn          = {1742-5468},
  month         = {May},
  number        = {5},
  pages         = {P05013},
  volume        = {2014},
  archiveprefix = {arXiv},
  doi           = {10.1088/1742-5468/2014/05/p05013},
  eprint        = {1401.3394},
  primaryclass  = {cond-mat.stat-mech},
  publisher     = {IOP Publishing},
}

@Article{PMGSc2014,
  author        = {Muratore-Ginanneschi, Paolo and Schwieger, Kay},
  journal       = {Physical Review E},
  title         = {{How nanomechanical systems can minimize dissipation}},
  year          = {2014},
  month         = {December},
  number        = {6},
  pages         = {060102(R)},
  volume        = {90},
  archiveprefix = {arXiv},
  doi           = {10.1103/PhysRevE.90.060102},
  eprint        = {1408.5298},
  primaryclass  = {cond-mat.stat-mech},
}

@Book{FraT2012,
	author    = {Theodore Frankel},
	publisher = {Cambridge University Press},
	title     = {{The geometry of physics: an introduction}},
	year      = {2012},
	edition   = {3rd},
	isbn      = {978-0-521-53927-2},
	pages     = {LXII, 686},
	url       = {http://www.cambridge.org/9780521833301},
	eisbn     = {978-1-139-15414-7},
}

@Article{PeCu2019,
	author    = {Gabriel Peyr{\'{e}} and Marco Cuturi},
	journal   = {Foundations and Trends in Machine Learning},
	title     = {Computational Optimal Transport: With Applications to Data Science},
	year      = {2019},
	doi       = {10.1561/2200000073},
	eprint    = {1803.00567},
	number    = {5-6},
	pages     = {355--607},
	publisher = {Now Publishers},
	url       = {https://optimaltransport.github.io/book/},
	volume    = {11},
}

@Article{MaNeWy2008,
	author        = {Christian Maes and Karel Neto\v{c}n\'y and Bram Wynants},
	journal       = {Markov Processes and Related Fields},
	title         = {On and beyond Entropy Production: {T}he Case of {M}arkov Jump Processes},
	year          = {2008},
	eprint        = {0709.4327},
	number        = {3},
	pages         = {445–464},
	url           = {http://arxiv.org/abs/0709.4327},
	volume        = {14},
	archiveprefix = {arXiv},
	primaryclass  = {cond-mat.stat-mech},
}

@Article{MGMePe2012,
	author        = {Paolo Muratore-Ginanneschi and Carlos Mej\'ia-Monasterio and Luca Peliti},
	journal       = {Journal of Statistical Physics},
	title         = {{Heat release by controlled continuous-time Markov jump processes}},
	year          = {2013},
	doi           = {10.1007/s10955-012-0676-6},
	eprint        = {1203.4062},
	number        = {1},
	pages         = {181-203},
	url           = {https://link.springer.com/article/10.1007/s10955-012-0676-6},
	volume        = {150},
	archiveprefix = {arXiv},
	month         = {January},
	primaryclass  = {cond-mat.stat-mech},
}

@Book{ZwaR2001,
	author    = {Robert Zwanzig},
	publisher = {Oxford University Press},
	title     = {{Nonequilibrium statistical mechanics}},
	year      = {2001},
	isbn      = {978-0-19-514018-7},
	pages     = {240},
	url       = {http://ukcatalogue.oup.com/product/9780195140187.do},
}

@Article{MaReMo2000,
	author          = {Christian Maes and Frank Redig and Annelies Van Moffaert},
	journal         = {Journal of Mathematical Physics},
	title           = {{On the definition of entropy production, via examples}},
	year            = {2000},
	doi             = {10.1063/1.533195},
	number          = {3},
	pages           = {1528--1554},
	volume          = {41},
	acknowledgement = {#ack-nhfb#},
	bibdate         = {Fri Oct 20 08:10:40 MDT 2000},
	bibsource       = {http://www.aip.org/ojs/jmp.html},
	coden           = {JMAPAQ},
	issn            = {0022-2488},
	month           = {March},
}

@Article{LeSp1999,
	author         = {Joel L. Lebowitz and Herbert Spohn},
	journal        = {Journal of Statistical Physics},
	title          = {{A Gallavotti-Cohen Type Symmetry in the Large Deviation Functional for Stochastic Dynamics}},
	year           = {1999},
	doi            = {10.1023/A:1004589714161},
	eprint         = {cond-mat/9811220},
	number         = {1},
	pages          = {333--365},
	url            = {http://arxiv.org/abs/cond-mat/9811220},
	volume         = {95},
	archiveprefix  = {arXiv},
	citeseerurl    = {http://www.ingentaconnect.com/content/klu/joss/1999/00000095/F0020001/00220489},
	month          = {March},
	oai2identifier = {cond-mat/9811220},
	primaryclass   = {cond-mat.stat-mech},
}

@Article{SekK1998,
	author    = {Sekimoto, Ken},
	journal   = {Progress of Theoretical Physics Supplement},
	title     = {{Langevin Equation and Thermodynamics}},
	year      = {1998},
	doi       = {10.1143/PTPS.130.17},
	pages     = {17-27},
	volume    = {130},
	adsurl    = {http://adsabs.harvard.edu/abs/1998PThPS.130...17S},
}

@Article{MeyP1982,
	author    = {Meyer, Paul-Andr\'e},
	journal   = {S\'eminaire de probabilit\'es de Strasbourg},
	title     = {{G\'eom\'etrie diff\'erentielle stochastique, II}},
	year      = {1982},
	pages     = {165-207},
	url       = {http://www.numdam.org/item?id=SPS_1982__S16__165_0},
	volume    = {S16},
	}
\bibliographystyle{unsrt} %base bibstyle abbrv.bst with eprint field
%\bibliography{/Users/paolo/RESEARCH/BIBTEX/jabref}{} 

\end{document}